\documentclass[usenatbib]{mnras}

\usepackage{amsmath, amssymb}
\usepackage{siunitx}
\usepackage{xfrac}
\usepackage{xcolor}
\usepackage{graphicx}
\usepackage{booktabs}
\usepackage{csquotes}
\usepackage{hyperref}   
\usepackage{orcidlink}
            
\hypersetup{pdfpagemode = {UseNone},
            pdftitle = {Planet composition},
            pdfcreator = {\LaTeX},
            pdfproducer = {pdfeTeX-0.\the\pdftexversion\pdftexrevision},
            pdfauthor = {AP, RAB},
            pdfsubject = {},
            pdfview = {FitH},
            pdfstartview = {FitH},
            colorlinks = {true},
            linkcolor = [rgb]{0,0.35,0.7},
            citecolor = [rgb]{0,0.35,0.7},
            filecolor = [rgb]{0.61,0,0},
            urlcolor = [rgb]{0.61,0,0},
           }
           
\usepackage{bm}

\newcommand{\prt}[2]{\frac{\partial #1}{\partial #2}}

\colorlet{darkgreen}{green!60!black}

\newcommand{\St}{\mathrm{St}}
\newcommand{\rh}{r_\mathrm{hill}}

\newcommand{\beq}{\begin{equation}}
\newcommand{\eeq}{\end{equation}}

\title[BOWIE-ALIGN: Modelling the compositions of disc- and high-e migrated gas giants]{BOWIE-ALIGN: How formation and migration histories of giant planets impact atmospheric compositions}

\author[Penzlin, Booth et al.]{Anna~B.T. Penzlin$^{{\orcidlink{0000-0002-8873-6826}},1}$\thanks{E-mail: a.penzlin@imperial.ac.uk}\thanks{Authors contributed equally.},
Richard~A. Booth$^{{\orcidlink{0000-0002-0364-937X}},2}$\footnotemark[2],
James Kirk$^{\orcidlink{0000-0002-4207-6615}1}$,
James E. Owen$^{\orcidlink{0000-0002-4856-7837}1}$,
E. Ahrer$^{\orcidlink{0000-0003-0973-8426}3}$,
\newauthor
Duncan A. Christie$^{\orcidlink{0000-0002-4997-0847},3}$,
Alastair B. Claringbold$^{\orcidlink{0000-0003-1309-5558},4,5}$,
Emma Esparza-Borges$^{\orcidlink{0000-0002-2341-3233},6,7}$,
M. L\'opez-Morales$^{{\orcidlink{0000-0003-3204-8183}},8}$,
\newauthor
N. J. Mayne$^{{\orcidlink{0000-0001-6707-4563}},9}$,
Mason McCormack$^{{\orcidlink{0000-0002-1463-9847}},10}$,
Annabella Meech$^{{\orcidlink{0000-0002-7500-7173}}8,11}$,
Vatsal Panwar$^{\orcidlink{0000-0002-2513-4465},4,5}$,
Diana Powell$^{\orcidlink{0000-0002-4250-0957},10}$,
\newauthor
Denis E. Sergeev$^{{\orcidlink{0000-0001-8832-5288}},9}$,
Jake Taylor$^{\orcidlink{0000-0003-4844-9838},11}$,
Peter J. Wheatley$^{{\orcidlink{0000-0003-1452-2240}},4,5}$,
Maria Zamyatina$^{{\orcidlink{0000-0002-9705-0535}},9}$,
\\
$^{1}$Astrophysics Group, Department of Physics, Imperial College London, Prince Consort Rd, London, SW7 2AZ, UK \\
$^{2}$School of Physics and Astronomy, University of Leeds, Leeds LS2 9JT, UK\\
$^{3}$Max Planck Institute for Astronomy (MPIA), K\"{o}nigstuhl 17, 69117 Heidelberg, Germany\\
$^{4}$Centre for Exoplanets and Habitability, University of Warwick, Gibbet Hill Road, Coventry CV4 7AL, UK\\
$^{5}$Department of Physics, University of Warwick, Gibbet Hill Road, Coventry CV4 7AL, UK\\
$^{6}$Instituto de Astrof\'isica de Canarias, San Crist\'obal de La Laguna, Tenerife E-38200, Spain \\
$^{7}$Departamento de Astrof\'isica, Universidad de La Laguna, San Cristóbal de La Laguna, Tenerife E-38200, Spain\\
$^{8}$Center for Astrophysics ${\rm \mid}$ Harvard {\rm \&} Smithsonian, 60 Garden St, Cambridge, MA 02138, USA\\
$^{9}$Department of Physics and Astronomy, Faculty of Environment, Science and Economy, University of Exeter, Exeter EX4 4QL, UK\\
$^{10}$Department of Astronomy and Astrophysics, University of Chicago, IL, 60657, USA\\
$^{11}$Department of Physics, University of Oxford, Keble Road, Oxford, OX1 3RH, UK\\
}

\date{Accepted XXX. Received YYY; in original form ZZZ}

\pubyear{2024}

\begin{document}
\label{firstpage}
\pagerange{\pageref{firstpage}--\pageref{lastpage}}
\maketitle

\begin{abstract} 
{
Hot Jupiters present a unique opportunity for measuring how planet formation history shapes present-day atmospheric composition. However, due to the myriad pathways influencing composition, a well-constructed sample of planets is needed to determine whether formation history can be accurately traced back from atmospheric composition.
To this end, the BOWIE-ALIGN survey will compare the compositions of 8 hot Jupiters around F stars, 4 with orbits aligned with the stellar rotation axis and 4 misaligned.
Using the alignment as an indicator for planets that underwent disc migration or high-eccentricity migration, one can determine whether migration history produces notable differences in composition between the two samples of planets. 
This paper describes the planet formation model that motivates our observing programme. Our model traces the accretion of chemical components from the gas and dust in the disc over a broad parameter space to create a full, unbiased model sample from which we can estimate the range of final atmospheric compositions.  
For high metallicity atmospheres ($O/H\geq10 \times$\,solar), the C/O ratios of aligned and misaligned planets diverge, with aligned planets having lower C/O ($<0.25$) due to the accretion of oxygen-rich silicates from the inner disc. However, silicates may rain out instead of releasing their oxygen into the atmosphere. This would significantly increase the C/O of aligned planets (C/O $>0.6$), inverting the trend between the aligned and misaligned planets. Nevertheless, by comparing statistically significant samples of aligned and misaligned planets, we expect atmospheric composition to constrain how planets form.
}
\end{abstract}

\begin{keywords}
Planet formation -- Protoplanetary discs
\end{keywords}

\section{Introduction}\label{sec:intro}

Thanks to ALMA and JWST, our ability to measure the chemical outcome of planet formation has increased significantly over the last decade. 
ALMA observations enable us to constrain the chemical composition of the protoplanetary environment \citep[e.g.][]{Cleeves2018,Miotello2019,MAPS,Bosman2021,Calahan2023} and JWST massively improved the sensitivity to the carbon and oxygen content of exoplanet atmospheres and discs \citep[e.g.,][]{JWST2023,Rustamkulov2023,Alderson2023,Ahrer2023,Feinstein2023, Tabone2023, Banzatti2023, Munoz-Romero2024, 2024MINDS}. 
It has long been suggested that, together, these advances may enable us to build empirical evidence as to how planets form by linking the composition of the planets to their building blocks \citep[e.g.,][]{Oberg2011, Madhusudhan2014}. 

However, the reliability of this approach remains unclear as numerous uncertain factors can affect both disc and exoplanet compositions. For example, gas-phase and grain surface reactions in protoplanetary discs may lead to significant changes in abundance ratios such as C/O over the Myr time-scales on which giant planets are thought to grow \citep[e.g.][]{Eistrup2016, 2018Eistrup, 2022Molliere}, and the extent to which this occurs needs to be critically tested through observations. Similarly, the amount of carbon contained in refractory carbon grains and the location where carbon grains are destroyed can also dramatically change the composition of the inner disc \citep[e.g.][]{Cridland2019cd, Bergin2023}. The planet formation models are also uncertain; for example, the relative importance of planetesimals and pebbles for growing giant planet cores remains debated \citep[see, e.g., the review by][]{Drazkowska2023}, with different assumptions leading to different compositions for both discs and planets \citep[e.g.][]{Madhusudhan2014, 2017Madhu, Booth2017, Danti2023}. Further, there are challenges linking the \emph{bulk} compositions of exoplanets predicted by formation models to the \emph{atmosphere} compositions measured by JWST, given that the solar system giants are known to host composition gradients \citep{Wahl2017,Vazan2018,Helled2024} resulting in atmospheric compositions that do not necessarily reflect the bulk composition \citep{FOnte2023,2024Muller,Calamari2024}. Also, the late accretion of, e.g., comets might further affect the planet's composition after formation \citep{2024SM}. 
In addition, the observation of the atmosphere reveals partial measurements that will shift due to the presents of clouds \citep{2016Helling} or the strength of atmospheric mixing and winds  \citep{2023Maria}.
While these effects can lead to complicated predictions that make it difficult to interpret the atmospheric compositions of individual exoplanets \citep[e.g.][]{Notsu2020}, some trends remain clear. For example, we may expect exoplanet C/O ratios to be correlated with disc metallicity \citep[e.g.][]{Cridland2019} and species such as sulphur, sodium, or silicon to trace the amount of solids accreted \citep{Kama2019, Turrini2021}. 

Given these complexities, it is natural to ask whether it is possible to identify a sample of exoplanets that: i) likely had different formation histories and ii) are relatively amenable to the characterisation of their atmospheric compositions. Such a sample can be used to test whether exoplanet atmosphere compositions correlate with their formation history and determine which planet formation aspects are most readily addressed through atmospheric compositions. The BOWIE+ (Bristol, Oxford, Warwick, Imperial, Exeter, +) collaboration's BOWIE-ALIGN survey (A spectral Light Investigation into hot gas Giant origiNs, \citep{2024survey}) seeks to achieve this by characterising the atmospheric composition of eight hot Jupiters with JWST, four with orbits aligned with their stellar spin axes and four that are misaligned \citep{2024survey}.

A planet's alignment relative to the stellar spin axis (its `obliquity') is a likely indicator of its migration history.
High obliquities, or strongly misaligned orbits, are hypothesised to be the result of high-eccentricity migration after the disc dispersal \citep[e.g.,][]{Rasio1996,Wu2011,2016Munoz} from orbits beyond $\sim1$\,au. Low obliquities, or spin-aligned orbits, are a natural outcome of migration through the protoplanetary disc to orbits $\leq0.1$\,au.
This is true if the planet orbits a star above the Kraft break where tidal realignment is thought to be inefficient and thus primordial obliquities are maintained \citep[e.g.,][]{Spalding2022}.
Hence, our JWST programme will refine and test model parameters and assumptions by observing planets with compelling independent evidence that they underwent different migration pathways and ended their growth in different parts of the disc.

The basic premise was already outlined by \citet{Madhusudhan2014}, who argued that high-eccentricity migrated (misaligned) hot Jupiters may be expected to have bulk compositions with higher C/O ratios than aligned hot Jupiters as a consequence of them accreting their solids and gas further from the star. This arises because more carbon-containing molecules are frozen out further from the star, leading to solids with higher C/O ratios. 

In the current work we present the theoretical groundwork to investigate and understand BOWIE-ALIGN programme observations \citep{2024survey} of aligned and misaligned hot giant planets around F-stars, exploring the many vagaries of planet formation. Our work explores what compositions are possible and what we expect to learn if  specific trends are observed in the BOWIE-ALIGN sample.

Section \ref{sec:model} describes the planet formation model and the choice and range of model parameters included in our investigation. The full set of models for possible planet evolutions and the resulting atmospheric C/O ratio is shown in Sec.~\ref{sec:full-sample}. In Section \ref{sec:assumption}, we discuss the effects of atmospheric mixing, silicate rain-out and the amount of carbon refractories. Section~\ref{sec:insight} relates the results of the model to the potential outcome of JWST observations. We summarise our conclusions in Section~\ref{sec:conclusions}.

\section{Planet formation model}\label{sec:model}

To understand the evolution in composition during planet formation, we build a simple model of a planet growing inside a gas and dust disc containing various molecular species.
With this model, we explore how planet composition depends on assumptions regarding the formation process and the uncertain disc environment, e.g., disc mass, temperature, dust, and planetesimal flux. 
The BOWIE-ALIGN programme planets orbit F-stars above the Kraft break that do not tidally align the planets.
Therefore,
this investigation focuses on F-star systems 
rather than solar analogues.
Our approach closely follows existing state-of-the-art planet evolution models \citep{2013Fortier,Booth2017,Schneider2021,Danti2023}. The model is designed to be simple enough to explore a large range of parameters, investigating the variety of possible outcomes for different growth scenarios of hot gas giants. In our model, we simplify the disc to a steady state within each simulation, while the range of different parameters cover different stages of the evolution of the disc at which the planets may form.
A detailed description of the model can be found in Appendix \ref{sec:planet}. 

The planet grows by accreting solids and gases from the surrounding disc.
We adapted the description of planet growth from \cite{2015Bitsch} for planet formation via pebble accretion to include a parameterised radially dependent planetesimal accretion based on the findings in \cite{2013Fortier}; a similar approach was taken by \citet{Danti2023}.
The planets accrete from the local environment, composed of dust and gas. For simplicity, the surface density of planetesimals is set to a constant multiple of local dust surface density. This planetesimal-to-dust mass fraction $f_{\rm pl}$ is varied between $10^{-5}-10^{0}$. 
Especially for planet growth in the inner disc where the pebble accretion isolation mass is lower, planetesimal accretion is the dominant contributor to solid accretion to the atmosphere.
Since we consider atmospheric growth of super-Earth protoplanets ($m_\mathrm{ini}=5~M_\oplus$) into giants, planet migration is treated using a simplified prescription dependent on the viscous speed, planet mass and disc density derived from fits to the results of the simulations by \cite{2015Durmann}.

We compute the planet's composition by tracking the composition of the solids and gas accreted. Our standard assumption is that all solids fully evaporate as they pass through the planet's envelope, mixing with the gas. However, since the planets can accrete a significant fraction of their envelopes at distances of several au from their host stars, the upper atmospheres of these planets can be quite cool. For this reason, it is possible that silicates may not evaporate until deeper in the atmosphere and, therefore, do not contribute to the atmosphere composition \citep[see, e.g.][]{Ormel2021}. For this reason, we also investigate a scenario in which silicates do not evaporate into the atmosphere but rain out so that they do not contribute to the oxygen abundance. Finally, we investigate the possibility that the atmospheres are not homogeneously mixed with a uniform composition by comparing our bulk abundances to those derived from the final 10\% or 1\% of accreted material. Such simple model assumptions change the final atmospheric composition more substantially than most system parameters.

The protoplanetary disc is modelled as a steady-state viscous disc with a Shakura-Sunyaev viscosity of $\alpha=10^{-3}$. 
We fix the $\alpha$ parameter since its effects are degenerate with the other parameters, such as disc mass and disc temperature.
We include dust dynamics by specifying a constant Stokes number, $St$, of the dust grains, which controls how fast the dust grains migrate towards the star. 
At low Stokes numbers ($St \ll \alpha$), dust and gas are well-coupled, moving together. 
The dust-to-gas mass flux ratio is set at the outer boundary, at which all molecular species but H$_2$ are in solid form. The steady state disc adjusts all molecular species to match the radially constant mass flux.
Thus, increasing the dust-to-gas flux increases the disc's overall metallicity.
The composition of the gas and solids in the disc is computed by solving for the adsorption and thermal desorption of molecules from the grain surfaces, as outlined in Appendix~\ref{sec:disc}, together with the advection and diffusion of the gas and ices, as in \citet{Booth2017, Booth2019} or \citet{Schneider2021}. We do not include any chemical conversion of one species into another, instead investigating the impact of chemical evolution by exploring different abundances of the key molecules. The abundances of each of the molecular species are chosen to produce a total composition equal to the \citet{Asplund2009} solar composition.

Once the composition is specified,  the model leaves us with six free parameters: the total disc mass parameterised by the mass accretion $\dot{M}_\text{g}$, the ratio of dust-to-gas fluxes in the disc $\dot{M}_\text{d}/ \dot{M}_\text{g}$, 
the dust's Stokes number over viscous coefficient St/$\alpha$, 
the fraction of mass in planetesimals relative to dust $f_\text{pl}$, the disc temperature, $T= T_0 (R[\mathrm{au}])^{-0.5}$, prescribed through the temperature at 1~au, $T_0$, and the initial distance of the planets to the star $r_\text{ini}$. All of these have been varied across the ranges listed in Tab.~\ref{tab:models}.

\begin{figure}
    \centering
    \includegraphics[width=\columnwidth]{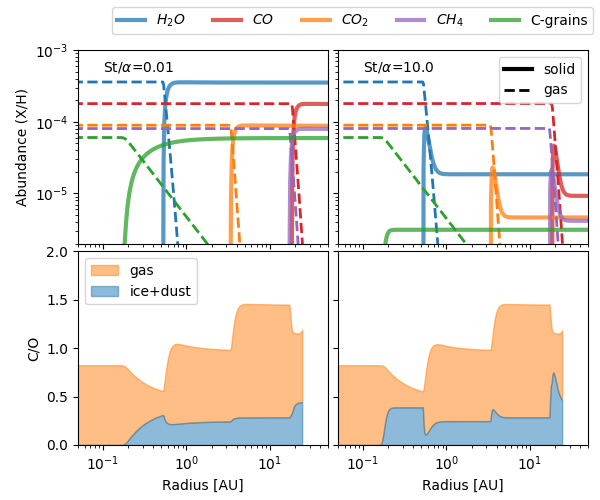}
    \caption{Molecule abundances throughout the disc in gas and ice (top) and the resulting C/O ratio (bottom). The left panels show a disc with nearly no dust drift. The right panels show the enhancements near the icelines of fast drifting dust. To achieve a constant flux in all discs, the level of solid composition to gas composition scales with the drift velocity.}
    \label{fig:disc}
\end{figure}

An example of the disc composition and resulting local C/O ratio is shown in Fig.~\ref{fig:disc} for two models with the same accretion rates of gas and solids but different Stokes numbers, $St$, and associated radial drift speeds. 
At the same $\dot{M}_\text{d}/ \dot{M}_\text{g}$, gas phase abundances of the different models are 
unaffected by $St$, except for very localised changes near the iceline as it is the total mass flux, not the speed of transport, that sets the composition. 
The dust-to-gas ratio and ice phase abundances are lower for higher $St$, due to the faster radial drift at fixed dust flux for increased Stokes numbers.
Fig.~\ref{fig:disc} also shows the familiar results that the C/O ratio of the gas and solids changes across ice lines. When radial drift is important, spikes in the ice-phase abundances appear close to ice lines because of the diffusion of gas-phase molecules back across the ice lines \citep{Oberg2016,Booth2017}. Since spikes in gas abundance near ice lines are a transient feature, they are not captured by our steady-state model (discussed in Sec.~\ref{sec:discuss}).

\begin{table}
    \caption{Sets of parameters. All combinations of varying parameters in one row have been modelled except for $r_\text{ini}$ which corresponds to the accretion rate $\dot{M}_\text{g}$-values. The remaining parameters are dust-to-gas flux $\dot{M}_\text{d}/ \dot{M}_\text{g}$, the Stokes over $\alpha$ as an indicator for the solid drift in the disc, the reference temperature at 1au $T_0$ and the planetesimal-to-dust mass fraction $f_\text{pl}$. Values with * are only realized with $M_\mathrm{gas} =10^{-8}$.
   }
    \label{tab:models}
    \centering
    \begin{tabular}{l|cc}
         Parameter & fiducial & range \\ \hline
         $\dot{M}_\text{g}$ [$M_\odot$/yr]& $10^{-8}$  & $10^{-9}$, $10^{-8}$,$10^{-7}$ \\
         $\dot{M}_\text{d}/ \dot{M}_\text{g}$  & 0.01 & 0.01, 0.05$^*$, 0.1$^*$ \\
         St/$\alpha$& 10 & 0.01, 1$^*$, 10 \\
         $r_\text{ini}$ [au]& 5.5-15.5 & 5.5-[12.5,15.5,20.5]\\
         $T_0$ [K]& 150 & [125,136,150 \\
         && 165,182,200] \\
         $f_\text{pl}$& 0.01 & $10^{-5}-10^{0}$\\
         
    \end{tabular}
\end{table}

\section{A zoo of modelled close-in giants}\label{sec:full-sample}

\begin{figure}
{\includegraphics[width=\columnwidth]{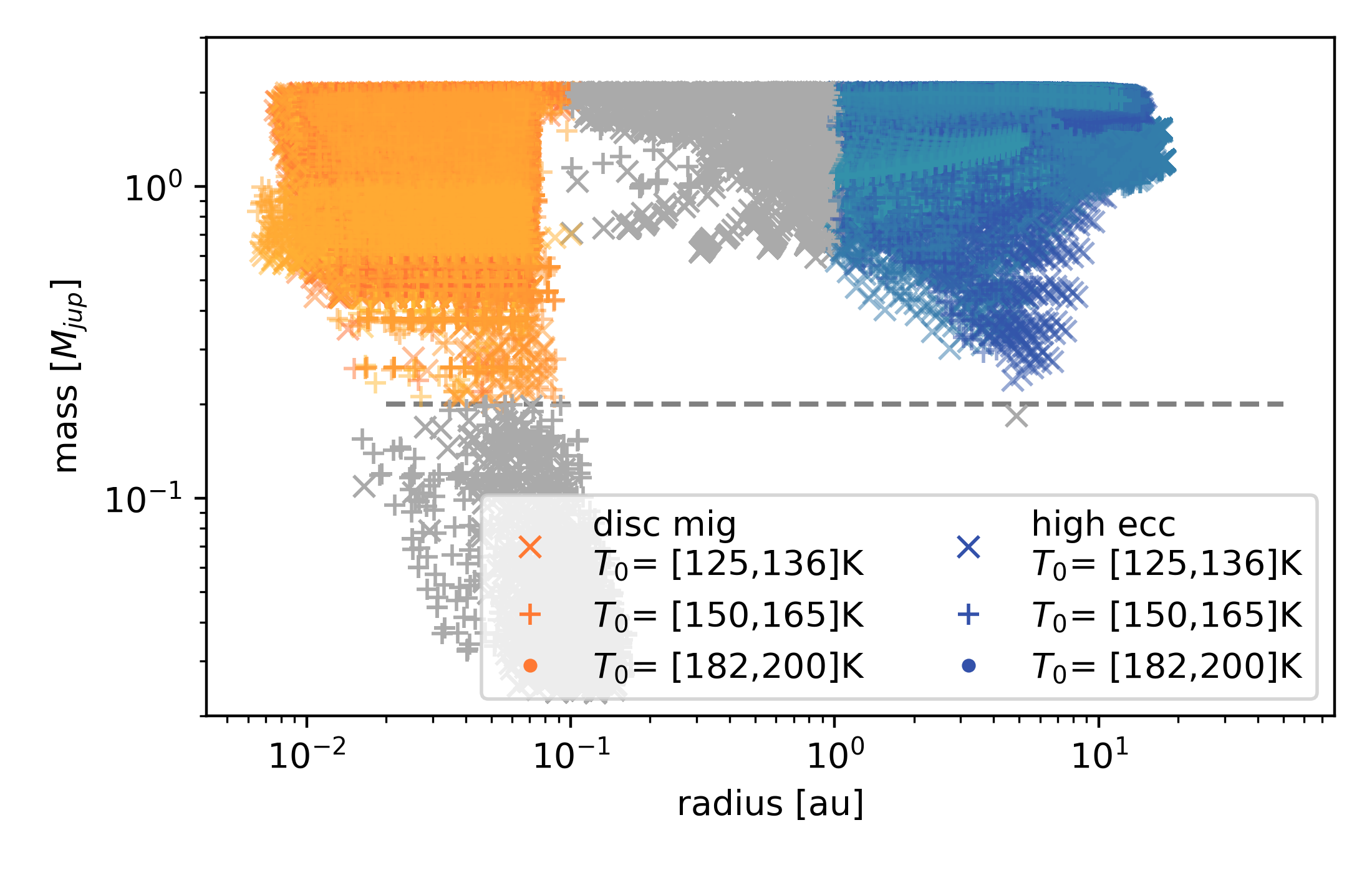}}
{\caption{The final masses and positions of planets from all 180000 runs. The marker symbol corresponds to the temperature ($T_0$) and the marker colour corresponds to the gas accretion, Stokes number and dust-to-gas ratio. Orange markers represent the disc migrated planets and cyan to blue markers represent the high-eccentricity migrated planets. The dashed line shows the lower mass cut-off of the sample. The planets in grey are excluded from the sample.}\label{fig:all_planets}}
\end{figure}

\begin{figure}
{\includegraphics[width=\columnwidth]{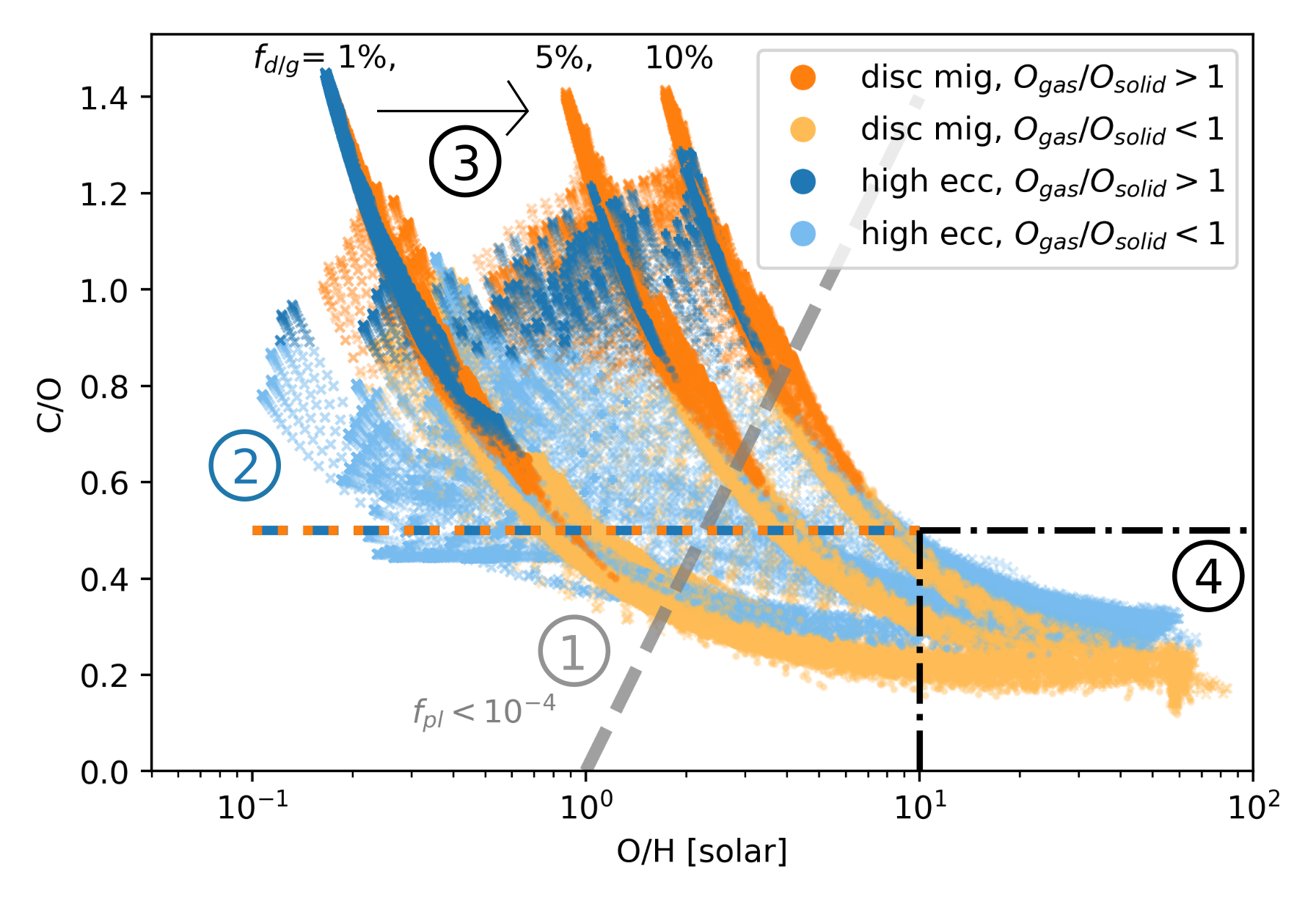}}
{\caption{The final atmosphere compositions from the full parameter range in Table~\ref{tab:models}. The marker symbol corresponds to different temperatures at 1~au ($ \times \rightarrow 125$~K, $ + \rightarrow 150$~K, $\circ \rightarrow 200$~K). Orange markers represent the disc migrated planets and blue markers represent the high-eccentricity migrated planets. Saturated markers show gas dominated accretion $\mathrm{O_{gas}}/\mathrm{O_{dust}}>1$ where the majority of oxygen in the planet is accreted through gas. Bright markers show dust dominated accretion $\mathrm{O_{gas}}/\mathrm{O_{dust}}<1$ where the majority of oxygen is accreted through dust. The three different tracks correspond to different dust-to-gas flux ratios and are labelled. The dashed grey line indicates the critical values for models with planetesimal accretion fraction less than $<10^{-4}$, all such models are confined to the left side of the line. The coloured dashed line indicates the minimum C/O ratio for gas dominated accretion ($\mathrm{O_{gas}}/\mathrm{O_{dust}}>1$).}\label{fig:all_set}}
\end{figure}

\begin{figure}
{\includegraphics[width=\columnwidth]{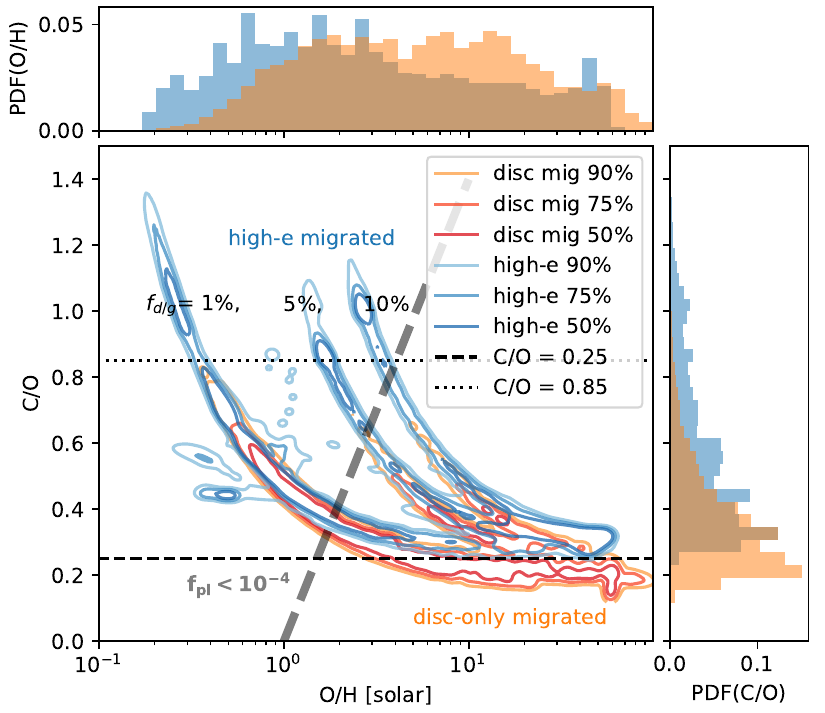}}
{\caption{Probability distribution of planets for all model parameters shown in Fig.~ \ref{fig:all_set}. The models have been given weights to produce samples with a uniform distribution of all input parameters on a linear scale. Thus gas dominated models are down-weighted compared to Figure \ref{fig:all_set}. Orange colours show disc-migrated planets and blue colours show high-eccentricity migrated planets. The colours from light to dark incorporate [90\%, 75\%, 50\%] of the total sample. The probability density function (PDF) of both the C/O and O/H ratio is displayed to the right and top of the figure.
Only planets that purely migrated through the disc can reach C/O ratios below 0.25 and planets with C/O above 0.85 are overwhelmingly more likely to have assembled beyond 1~au if any planetesimal accretion occurs.}\label{fig:weight_dist}}
\end{figure}

We begin by studying the compositions of planets formed in `standard' models of planet formation. Here, we vary the disc accretion rate (and therefore mass), its temperature, and the ratio $\dot{M}_{\rm d}/\dot{M}_g$. This latter parameter covers a range of scenarios from the efficient pebble drift used in typical pebble accretion models ($\dot{M}_{\rm d}/\dot{M}_\mathrm{g}\sim 0.1 $, c.f. \citealt{Booth2017}; \citealt{Booth2020}; \citealt{Danti2023}) to standard planetesimal accretion models without any enhancement ($\dot{M}_{\rm d}/\dot{M}_\mathrm{g}\sim 0.01 $). We also vary the initial formation location of the planets and the amount of planetesimals in the disc, $f_{\rm pl}$, to produce a range of planets with different metallicities\footnote{A single metallicity is not a well-defined concept for a general composition. Here, we will use the oxygen abundance relative to solar as a proxy for metallicity.} and accretion histories. In total, we simulated $1.8\times10^5$ planets across the full parameter range. 

To define our `disc migration' and `high-eccentricity migration' hot Jupiters we consider only planets with final masses $ 0.2 ~M_\mathrm{Jup} < m_\text{p} < 2~ M_\mathrm{Jup}$ to only study gas giants comparable to those observed in the BOWIE-ALIGN programme. For the disc-migrated planets, we consider only those that reach separations $<0.1$~au by the end of the simulation (73812 planets, orange markers, Fig.~\ref{fig:all_planets}) and the high-eccentricity migrated planets are considered as those that have final orbits $>1$~au (60545 planets, blue markers) \citep{2016Munoz}. We exclude the remaining planets from the analysis which are are too small or fall in between these separations(shown in Fig. 2 as grey symbols).
The model parameters are not distinguishable in the mass-distance plot in Fig.~\ref{fig:all_planets}, which highlights why an additional source of information is useful to constrain models.

The C/O and metallicity (explicitly defined as [O/H]), displayed in Fig.~\ref{fig:all_set} for all the simulations, cover a large parameter space, but it reveals 4 key behaviours:
\begin{itemize}
    \item[(1)] Planets that formed with only small amounts of solids accreted through planetesimals ($f_\text{pl}<10^{-4}$) remain below $5\times$ solar metallicity and above C/O$\gtrsim0.5$ (marked by the dashed grey line), consistent with previous results from pebble accretion models \citep{Booth2017, Schneider2021}.
    \item[(2)] Planet that are predominately enriched by gas accretion ($\mathrm{O_{gas}}/\mathrm{O_{dust}}>1$, saturated colours) have high C/O ratios of C/O$> 0.5$ throughout, due to methane and CO.
    \item[(3)] The high dust-to-gas mass flux ($f_{d/g}=10\%$) through the disc associated with efficient radial drift and pebble accretion causes all planets to be more metal-rich. This is particularly significant for high C/O ratios, where the majority of the metals were accreted in the form of gas (see \autoref{fig:disc}). The impact is less significant for planets with [O/H]$>10$ since these planets accrete most of their metals through planetesimals.
    \item[(4)] Above $10\times$ solar metalicity, the population of planets splits at C/O $\approx0.25$, with the disc-migrated planets (aligned/orange) being consistently more carbon depleted than the high-eccentricity migration (misaligned/blue) planets.
\end{itemize}

These results are broadly consistent with previous formation models. For example, the parameterised models of planetesimal accretion by \citet{Madhusudhan2014} predicted that the aligned (disc migration) planets should have lower C/O ratios than the misaligned (high-eccentricity migration) planets when [O/H] is super-solar, which is what we see here. Similarly, \citet{Cridland2019} suggested that in planetesimal accretion scenarios, C/O ratio and metallicity should be anti-correlated, while \citet{Booth2017} and \citet{Schneider2021} showed that planets that formed by pebble accretion without any enrichment of solids by planetesimals will have high C/O ratios since their metals predominantly come from the accretion of gas. These models predicted that moderately super-solar metallicities might be possible due to efficient radial drift, which we also see in our models.

The difference in C/O ratio between the disc migration and high-eccentricity migration planets arises primarily due to the change in the composition of the solids across the ice lines. Particularly critical to the low C/O ratio of the disc migration planets is the accretion of solids inside $\sim0.2$~au where the carbon grains are destroyed and, therefore, the planetesimals are then made of essentially pure silicates, which are carbon-free but contain oxygen. Since the positions of the ice lines are temperature dependent, warmer discs contribute to larger differences between the populations as the carbon destruction line (sometimes called the `soot line') is further out. 
In the regime of metal-poor planets, the small separations between icelines leads to a larger scatter in the abundance of the accreted gas which can be seen in the crosses in Fig.~\ref{fig:all_set}.
The different temperature models, thereby, blur the line between the populations of aligned and misaligned planets. Appendix \ref{sec:parameters} discusses the free parameters' effects, including temperature, in more detail.

\cite{2023Bitsch} recently proposed another pathway to increase the accretion of solids through the accretion of small dust grains inside of the water ice line. While we have not specifically modelled this process, the mechanism would be indistinguishable from our high metallicity planets formed through accreting planetesimals in the inner disc. 

In Fig.~\ref{fig:weight_dist}, all the models are combined into a sample density that now sampled with a uniform distribution of planetesimal fraction values.
Here, we find two clear separation lines: only planets that purely migrated through the disc can reach C/O ratios below 0.25; planets with C/O above 0.85 are overwhelmingly more likely to have assembled beyond 1~au if significant planetesimal accretion occurs.
Generally, planets assembled beyond 1~au are more metal-poor and have a higher C/O ratio, while fully migrated planets are more likely to be more metal-rich.

This zoo of potential outcomes in C/O and metallicity resulting from different formation scenarios presents a worst-case, unconstrained parameter space with which to compare to our observational programme. Here, there is substantial overlap between the aligned and misaligned planets populations, driven by the scatter in the underlying disc properties and assumptions of the planet formation model.
The models are designed to include the full range of uncertainty that exists from observational and theoretical constraints, however, it is likely that the astrophysical process of planet formation in our sample is more similar than this wide range of scenarios. Hence,
the C/O ratio and metallicity distributions of each sample will be narrower. In such a distribution, the aligned and misaligned populations are more clearly separated.
For reference, the detailed effects of individual parameters are further discussed in the Appendix.~\ref{sec:Temp}.

\section{Factors affecting planet compositions}\label{sec:assumption}

As the distribution of models in Sec.~\ref{sec:full-sample} explores the range of disc physics, the composition of the planet is also strongly affected by simple assumptions that change the chemistry or the accretion of different species.
Here, we explore the impact of common assumptions on the predicted atmospheric abundances.

First, we explore the fate of the dust grains. The bulk abundances of the giant planet envelopes are usually calculated assuming that dust grains are destroyed and the carbon and oxygen they contain are released into the atmosphere because the planet's envelopes are much hotter than the disc \citep[e.g.][]{Brouwers2018}. While this is likely a reasonable assumption for the carbon grains, which are likely destroyed at temperatures above $500\,{\rm K}$ \citep{2017Gail, Li2021}, silicates are more robust and likely only sublimate deep in the envelope.
Second, the models predict the bulk abundance of the gaseous envelope by adding up all of the material that enters it, and implicitly assume that the planet is well-mixed by assuming that atmospheric abundances and the bulk envelope abundances are the same. This is, however, doubtful as Jupiter and Saturn are known to have interior composition gradients \citep[e.g.][]{Wahl2017}.
Third, the gas-phase and solid carbon abundances in protoplanetary discs are poorly known. ALMA and Herschel observations show gas phase carbon abundances are depleted \citep{BErgin2013,McClure2016,Trapman2017,Zhang2021}, with freeze-out of CO onto cold grains \citep[e.g.][]{Xu2017,Powell2022} and chemical conversion of CO to less volatile species \citep[e.g.][]{Bosman2018} both being plausible explanations. As a result, the C/O ratio of solids close to the star is poorly known, potentially affecting the planet's composition.
The model allows us to study the impact of the unknown refractory carbon abundance and the depletion of CO through chemical reactions on the surface of grains. 
We investigate the impacts of these assumptions in the following sub-sections. 

\subsection{Inversion of the C/O trend through silicate condensation}\label{sec:noSi}

Silicates and other minerals carry a significant amount of bound oxygen transported in the disc. In our model, we consider MgFeSiO$_4$ as a rough composition for the dust. The exact dust composition is not critical and is chosen to follow the rough stoichiometry of two oxygen atoms per Si atom and one per Mg and Fe \citep[e.g.][]{Lodders2003,Woitke2018}. Silicates have much higher sublimation temperatures ($\sim 1500$~K) than carbon grains ($\sim 500$~K) and, as a result, are much more prone to condensing out of the planets' atmospheres. Recently, \citet{Powell2024} used 2D microphysical cloud models to show that silicates condense out of the upper atmospheres of hot Jupiters with equilibrium temperatures below about 1700~K. Since the hot Jupiters migrate from beyond the ice line towards their host stars during their formation histories, a significant amount of silicates could rain out, while the more volatile carbon-rich ices and organics can easily evaporate during accretion. The detection of SiO$_2$ clouds in the atmosphere of WASP-17b \citep{Grant2023} suggests that either silicate condensation during formation is incomplete or that, subsequently, the envelopes become well mixed. As a result, the degree of silicate loss from the atmosphere is uncertain. In light of these uncertainties, we compare our estimates of the bulk composition in the fiducial models with another set of models representing the opposite extreme - in which all of the silicates are assumed to condense out and not contribute to the atmospheric composition.

\begin{figure}
{\includegraphics[scale=0.55]{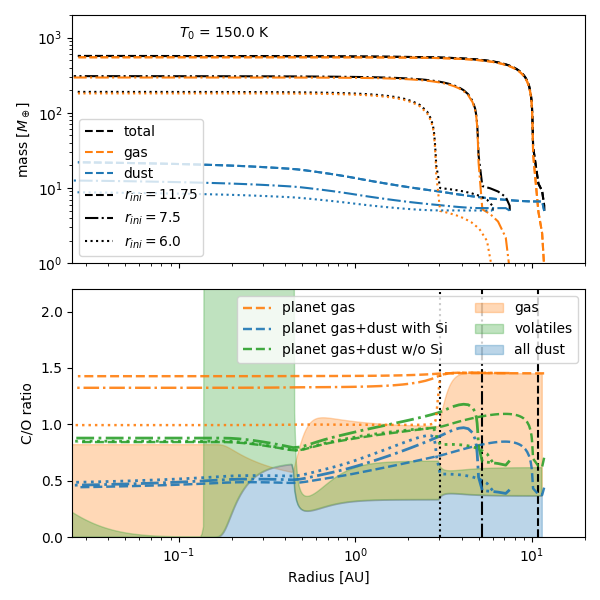}}
{\caption{The evolution in mass and C/O of three different test planets as a function of radius (semi-major axis) for a disc temperature of 150\,K. Top panel: the evolution of the gas mass (orange), dust mass (blue) and total mass (black) of three planets, initiated at radii of 6.0\,au (dotted), 7.5\,au (dashed-dotted) and 11.75\,au (dashed). The sharp increase in the gas and total mass indicates the location of runaway gas accretion. Bottom panel: The C/O ratio of the planets' gas (orange lines), gas and dust with silicates (blue lines), and gas and dust without silicates (green lines). The shaded regions indicate the C/O of the background disc components. The vertical black lines show the locations of the onset of runaway gas accretion for the three planets.}\label{fig:ex_Si}}
\end{figure}

Fig.~\ref{fig:ex_Si} shows how the C/O ratio depends on the contribution from silicates and volatiles. Considering only the frozen-out volatiles in the dust composition already leads to a much higher C/O ratio near a level of $\sim0.6$ in the outer disc, compared with values $\leq0.4$ when silicates are included.
When neglecting the silicates, the C/O ratio of the solids rises sharply inside the water ice line ($<0.5$~au) as only refractory carbon grains remain solid. With silicates included, the rise in C/O is limited to $\sim 0.6$.
Hence, if silicates condense out of the planet's atmosphere, the atmosphere will effectively be enriched by solids with C/O ratios $>0.6$, while atmospheres in which the silicates do not condense out would be more oxygen-rich with lower C/O.

When planets grow in the inner disc, they become more enriched in metals than planets growing for an equivalent time or to an equivalent mass in the outer disc. 
This is a result of the difference in both gas and dust accretion: At close-in orbits, the Hill spheres that the planets accrete gas from shrink, reducing the gas accretion, while planetesimal collisions become more likely on the faster orbits in a less vertically extended disc.
This is evident in the late accretion history of gas and solids in the top panel of Fig.~\ref{fig:ex_Si}. 
If minerals are considered, this leads to oxygen enrichment through oxygen-rich silicates. However, if minerals only contribute to the core, this would lead to a final phase of organic carbon accretion increasing the overall C/O ratio towards the level seen in the outer disc.

\begin{figure}
{\includegraphics[scale=0.6]{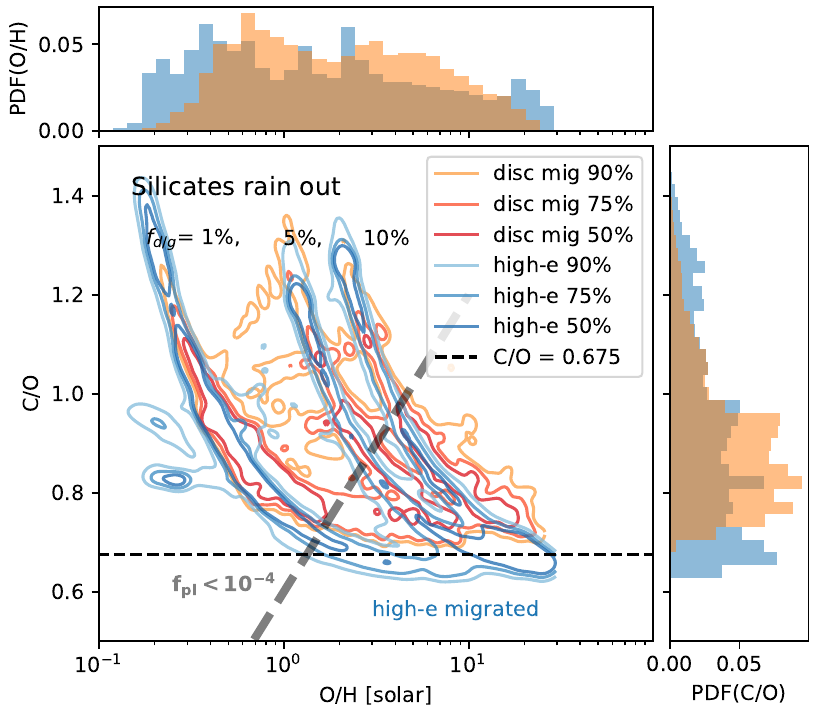}}
{\caption{The impacts of silicates on the planetary C/O and O/H. If O-bearing silicates are not evaporated into the atmosphere, the C/O remains high for both disc and high-e migrated planets ($>0.8$). If late enrichment by organic carbon species is considered, this can drive up the C/O of the disc-migrated planets, reversing the trend seen previously.}
\label{fig:weight_noSi}}
\end{figure}

When silicates rain out from the planets' atmospheres, the C/O ratio of the planets is higher and, more significantly, the difference in C/O between the aligned and misaligned planets is reversed for super-solar metallicities (\autoref{fig:weight_noSi}). This arises because the carbon-free planetesimals inside the carbon destruction front no longer contribute to the planets' composition. The difference in composition between the two samples of planets is now primarily driven by the accretion of the carbon-rich planetesimals between the water ice line and the carbon grain destruction front. This shows that understanding silicate condensation during planet formation and evolution will be essential to enable the C/O ratio (or indeed the S/O, N/O, or any ratio involving oxygen) to trace the formation history of any individual planet.

\subsection{Carbon carriers} \label{sec:carbon}

The main carbon carriers in protoplanetary discs are not well known. ALMA observations show that the gas in protoplanetary discs is typically depleted in carbon \citep[e.g.][]{Zhang2021}, which means that ices are critical for determining the main carbon and oxygen carriers, and the composition of ices is also not well known. Most studies of planet formation have chosen their compositions following the approach of either \citet{Oberg2011}, who used observations of the interstellar medium (ISM) as a proxy for protoplanetary disc compositions; \citet{Madhusudhan2014}, who based their model on equilibrium chemistry calculations \citep{Woitke2009}; or used estimates from the solar system \citep[e.g.][]{Pollack1994}. 

For our default abundances, we follow \citet{Oberg2019}, which are based on comets and the dense ISM. The model includes CO, CO$_2$, volatile organic carbon (for which we use CH$_4$ as a proxy), and refractory carbon grains. Roughly 40\% of the carbon is assumed to be refractory by default, in good agreement with observations of the ISM \citep{Mishra2015}, 15\% of the carbon in volatile organics and the remaining 45\% in CO and CO$_2$. For the exact values of abundances used, see \autoref{tab:chem_abund}.

\begin{figure}
{\includegraphics[width=\columnwidth]{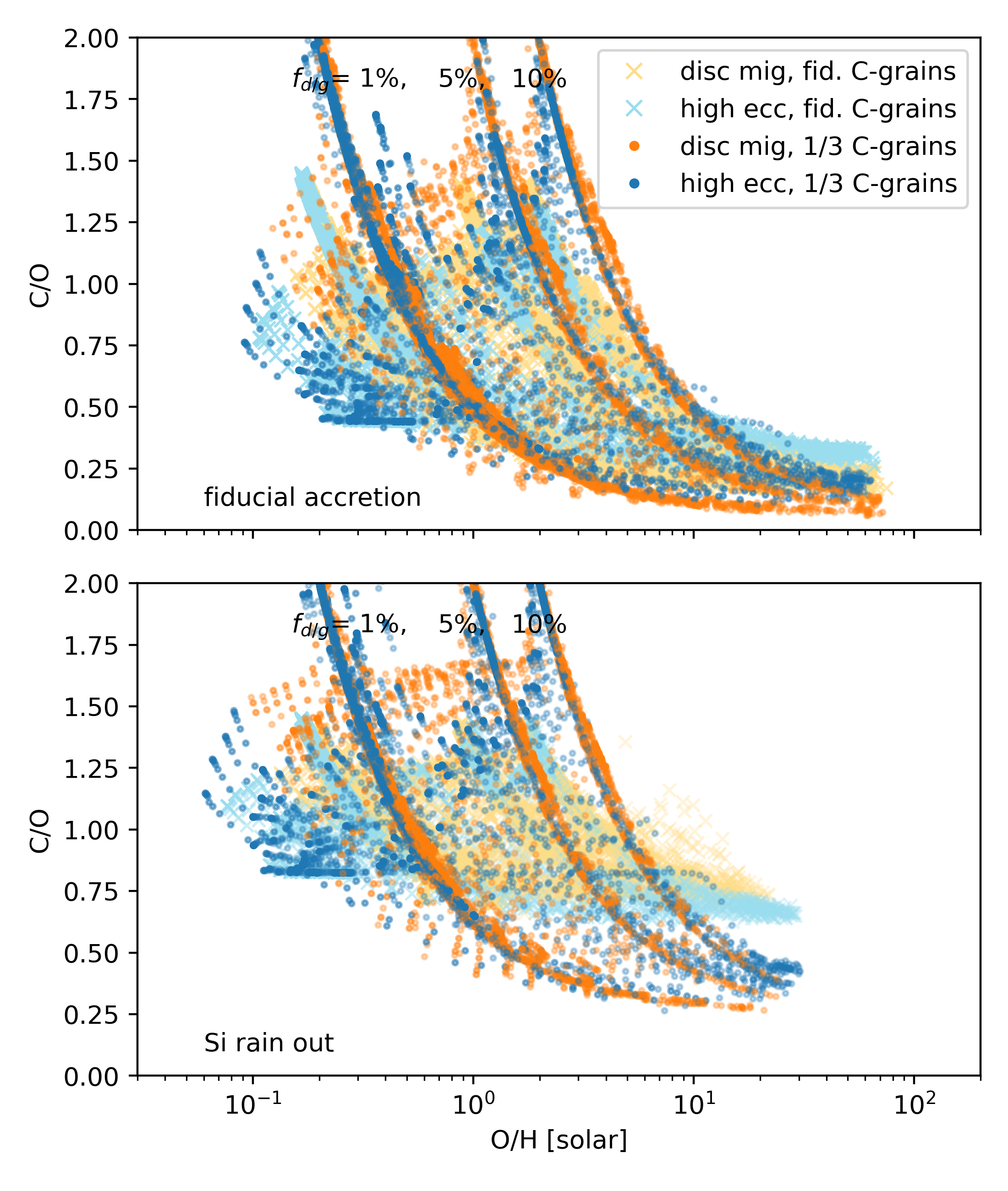}}
{\caption{The atmosphere compositions for high and low carbon refractory abundances.
Light orange and blue (x) indicate models with $75\%$ of organic carbon in refractory grains, while /new{saturated orange and blue (o)} indicate $25\%$. In the top panel, silicates enrich the atmosphere in oxygen. In the bottom panel, silicates do not contribute to the atmosphere composition.}
\label{fig:C_set}}
\end{figure}

As demonstrated in Section~\ref{sec:noSi}, the refractory carbon plays a significant role in the composition of the planets.
As a simple test of the importance of refractory carbon abundance, we swap the mass ratio of carbon in CH$_4$ and the refractory carbon grains from $1/3$ to $3/1$.
Fig.~\ref{fig:C_set} shows how the reduction in carbon grains leads to a larger range of C/O ratios in the overall distribution of models. This happens because the dominating contribution of carbon shifts from refractories to volatiles. 

Planets that are metal-poor due to little accretion of solids can be carbon-rich because the gas is carbon-rich, reaching C/O ratios above 2 between the CO$_2$ and CH$_4$ ice lines. Conversely, the smaller amounts of carbon in the refractory grains lead to a lower C/O ratio in the metal-rich planets. The change in the distribution of compositions is even more dramatic if silicates condense out (in the bottom panel Fig.~\ref{fig:C_set}). With the significant reduction of the carbon refractories in the inner disc $<1$~au, the relevant final accretion shifts to ice lines beyond the composition of the planet inside and outside 1~au, become degenerately similar and much more carbon-depleted. 

\begin{figure}
{\includegraphics[width=\columnwidth]{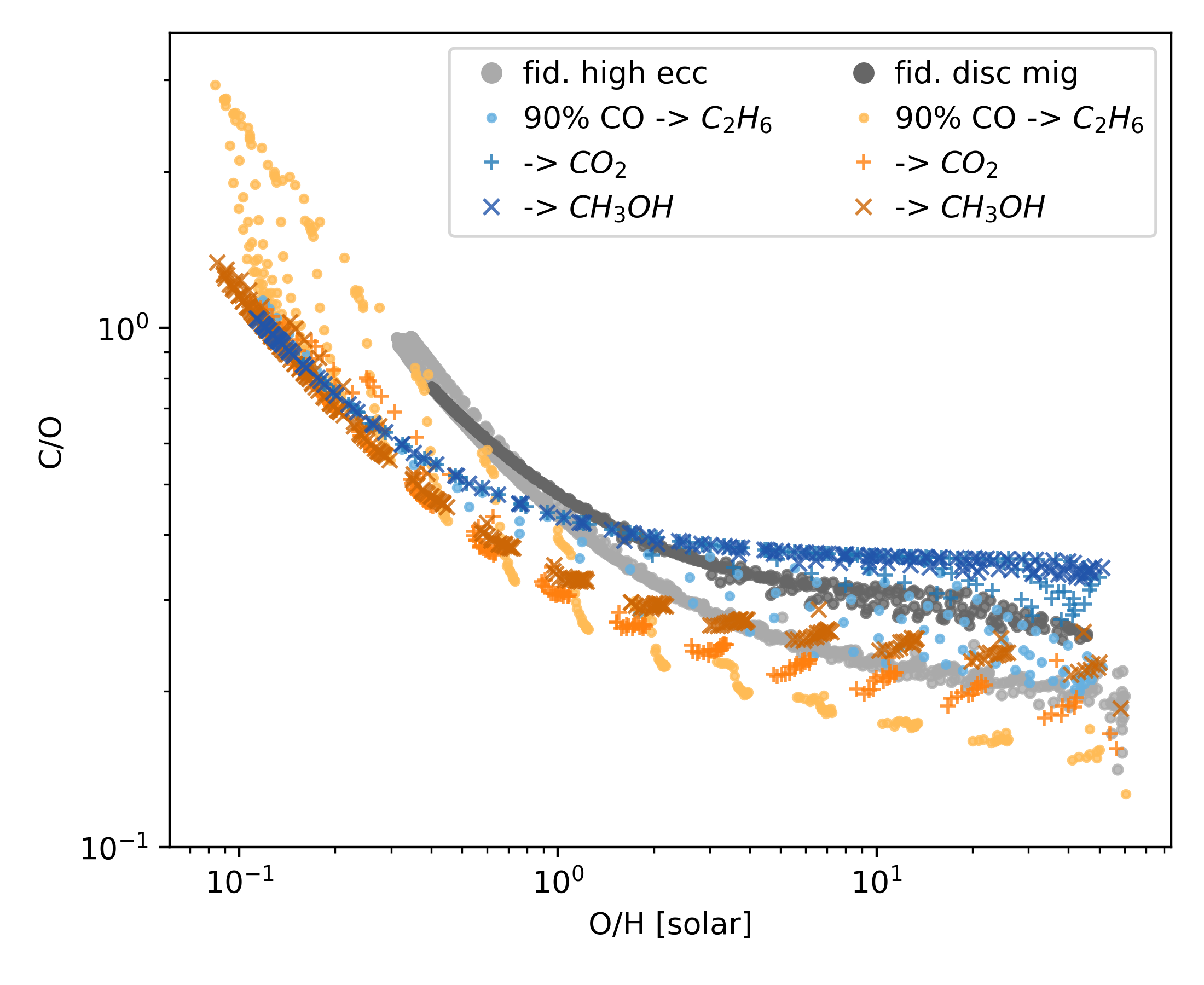}}
{\caption{Planet models for depleting CO into other species. Light ($ \cdot $) mark depletion into CH$_4$, saturated ($+$) mark CO$_2$, and dark ($\times$) mark methanol. Blue symbols are high eccentricity migrated planets and orange symbols disc migrated planets. Grey colours are the fiducial models for comparison.}\label{fig:depl}}
\end{figure}

In addition to the uncertain abundance of refractory carbon, the abundance of CO in protoplanetary discs is also uncertain. There is significant evidence that CO abundance in discs is depleted in the warm upper layers of discs probed by ALMA \citep[e.g][]{Bosman2021} and the extent to which this can be explained by the freeze-out of CO onto ices in the disc mid-plane alone is debated \citep[e.g.][]{Bosman2021_COdep, Powell2022,vanClepper2022}. Grain surface chemistry has been suggested to play an important role by converting CO to C$_2$H$_6$, CO$_2$, or CH$_3$OH, for example \citep{Bosman2018}. To understand how this can affect the composition of hot Jupiters, we explored what happens when $90\%$ of the CO is converted into one of these molecules (\autoref{fig:depl}). We adjusted the water abundance accordingly to ensure the correct (solar) oxygen abundance. 

Changing the CO abundance can significantly change the disc's metallicity because CO is the dominant source of carbon and oxygen in the region outside the CO$_2$ ice line ($\sim 3~{\rm au}$). Conversely, carbon species such as C$_2$H$_6$, CH$_3$OH, CO$_2$ evaporate at various ice lines and can be in solid form where planets accrete most of their gas. The lower gas-phase carbon and oxygen abundances due to the missing CO lead to generally less metal-rich gas and much larger shifts in the C/O ratio at these ice lines. 

The impact of CO depletion on the planet composition is presented in \autoref{fig:depl}, which shows the results of the fiducial model compared against equivalent models with CO depletion.
Planets that are not substantially enriched by planetesimals have a lower metallicity than in the fiducial models.
The biggest change in the C/O ratio occurs for the conversion of CO to C$_2$H$_6$ because it results in nearly all of the oxygen content being frozen out in water ice. This results in planets that have very high C/O ratios (up to 3) when their accretion is dominated by gases, if their runaway accretion happens inside the C$_2$H$_6$ ice line, which can happen for the disc-migrated planet population.
If the composition is instead dominated by solids, the increase of water ices leads to a reduction in C/O ratio for all planets.
Following the same logic, the conversion of CO to CO$_2$ or CH$_3$OH produces a smaller effect since these molecules contain carbon and oxygen.

While each of these depletion mechanisms can make a significant impact on the planet's composition, we find that the two different populations of planets maintain different compositions. However, the exact location and separation of the two samples shift up or down, for the disc migrated planets between C/O=0.15 for ethane to 0.3 for carbon refractories, and for the high eccentricity migration planets between C/O$>0.2$ for ethane to 0.4 for all other processes.
Thereby, carbon depletion adds an additional uncertainty to the model. 

If chemistry and planet formation processes are similar across all discs, the BOWIE-ALIGN programme is likely to find the differences between the two populations in each case. However, if the disc chemistry varies dramatically from system-to-system then atmospheric observations of a large number of planets may be needed to see the trends. The level of system-to-system variation will start to be addressed by the upcoming ALMA large programme ``The Disk Exoplanet C/Onnection'' (DECO ALMA large programme 2022.1.00875.L, PI: Ilse Cleeves / University of Virginia), after which it will be possible to assess the impact on the different planet populations more rigorously. Ultimately, the BOWIE-ALIGN programme will lay the groundwork for statistical constraints on planet formation derived from the chemical characterization of giant exoplanets.

\subsection{Interior mixing}\label{sec:fin10}

We explore the impact of incomplete mixing in the planet's envelope by comparing the bulk envelope abundances with the abundances of the final 10\% or 1\% of the material accreted.

\begin{figure}
{\includegraphics[width=\columnwidth]{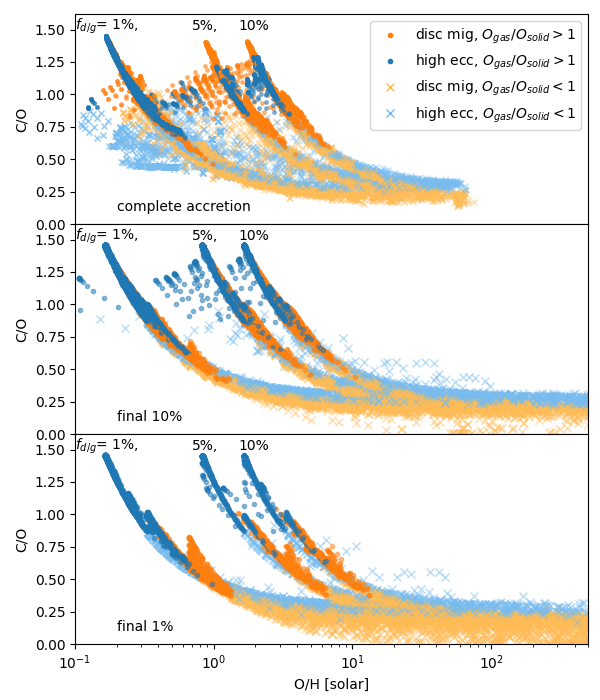}}
{\caption{The final atmosphere compositions when counting all, only the final 10\,\% and final 1\,\% of accreted material. The marker symbols and colours are the same as in Fig.~\ref{fig:all_set}.}\label{fig:fin_set}}
\end{figure}

An important factor in the composition of the final 10\% or 1\% of the material accreted is that gas accretion slows down once the planet becomes sufficiently massive to open a gap in the disc. At this point, pebble accretion might stop. While the gas accretion is reduced with growing mass and smaller orbits, the planetesimal accretion probability increases.
We can split the final accretion into 2 scenarios: accretion dominated by planetesimals and accretion dominated by gas. To separate the two scenarios, we look at the contribution of oxygen for gas relative to dust in the final accretion in Fig.~\ref{fig:fin_set}.

For planet models where the final oxygen accretion is dominated by solids, the significant solid or planetesimal accretion shifts the unmixed atmosphere towards lower C/O ratios and higher metallicities, as Fig.~\ref{fig:fin_set} shows.

The fully disc migrated planet sample accretes solids at the end of its formation that drive the planets' composition towards very low C/O ratios especially through silicates (see Fig.~\ref{fig:disc}). In comparison, the misaligned sample does not reach the region where solids have very low C/O ratios, and as a result, its composition is less affected by inefficient mixing. Therefore, incomplete mixing results in a larger difference in the C/O ratios for the two samples if solid accretion dominates, with the split remaining at C/O$\approx0.25$.

If gas accretion dominates and the majority of oxygen is accreted from gas, we find that planets are restricted to metallicities below the dust-to-gas flux (O/H$\leq \dot{M}_\mathrm{d}/\dot{M}_\mathrm{g} \times 100$) and C/O ratio above 0.5 in general.

Since more ice species sublimate close to the star, the gas species in the disc increase with decreasing separation, while the C/O ratio typically decreases (\autoref{fig:disc}). As a result, the final gas accreted is more metal-rich and has a lower C/O ratio for the disc migration planets, than the high-eccentricity migration case. 

If the atmosphere is less mixed, the C/O ratio of different paths of migration split near 0.75--1. All planets that only migrated through the disc have C/O$<1$. As Figure \ref{fig:disc} shows, the C/O ratio of gas reaches values as low as 0.5 at the water iceline and only rise slowly up to 0.8 inside the water ice line due to C-grain evaporation.
Planets that stalled their migration beyond $1$~au have C/O$>0.6$.
They accrete from a regime of the disc gas that is dominated by CO and CH$_4$ and potentially CO$_2$ for the further-in planets. The C/O ratio never drops below 1 in this regime.

This means that, in the most extreme case of very little mixing within the atmosphere (bottom panel in Fig.~\ref{fig:fin_set}), there is also a clear difference between the compositions of the aligned and misaligned planets, even without significant planetesimal accretion. Conversely, when mixing is efficient the late accretion of gas near the star contributes too little to the total to produce substantial differences in the populations. 
This implies that the BOWIE-ALIGN programme may be able to constrain the degree of mixing within the planet envelopes if it turns out that the planets all have metallicities below about $5\times$ solar and super-solar C/O ratios.

\subsection{Other considerations}\label{sec:discuss}

Our planet formation and disc composition models are necessarily simplified. 
The simplifications limited the modelling choices we considered. Even after exploring model assumptions and uncertain parameters, we did not consider other factors that can impact the difference between the aligned and misaligned hot Jupiters.

One important factor is the disc chemistry, as well as assumed composition -- the chemical conversion of one species to another -- which can affect the composition of the disc. \citet{Booth2019} argue that the transport processes considered in this act more quickly than chemical processes, but studies have shown that chemistry significantly changes the composition in certain parts of the disc. For example, \cite{2018Eistrup} showed that gas phase C/O ratio can decrease over several Myr due to the formation of O$_2$, particularly in the region around 5~au, where many of our planets accrete a substantial portion of their gas. If significant, this would decrease the difference in C/O between the super- and sub-solar metallicity planets, potentially reducing the difference between the compositions of the aligned and misaligned planets.
Another possibility is the removal of carbon-rich gases from the inner disc through internal photo-evaporation over the discs lifetime \citep{2024Lienert}, which would enhance the difference in the composition between aligned and misaligned planets even from reduced solid accretion.

In this work, we have also considered steady (time-independent) disc models. The steady condition considerably reduces the computational time and allows us to run a full range of scenarios that, through the considered different disc properties, can capture stages of the evolving disc within the complete sample.
These steady disc models do not produce spikes in the gas phase abundance of molecules inside the ice lines \citep[e.g][]{Oberg2016,Booth2017}, because this is a transient effect caused by the increase in $\dot{M}_\text{d}/ \dot{M}_\text{g}$ over time associated with grain growth. Given a sufficiently long time, the transient gas spikes seen in time-dependent models disappear as the molecules are carried inwards by gas accretion, leading to a uniformly increased abundance inside the ice line instead \citep[see, e.g.,][]{Booth2017}. Rather than using a time-dependent model, which would force us to specify when the planets form and how the dust evolves (by using the \citealt{Birnstiel2012} model, for example), we simply vary  $\dot{M}_\text{d}/ \dot{M}_\text{g}$ to allow for an enhancement of the molecular abundances arising from radial drift. The main difference between time-dependent models and a comparable steady-state model would be a somewhat increased scatter in the C/O ratio of the planets that are not substantially enriched by solids, owing to the larger range of disc gas compositions arising from the `spikes'.

A second significant factor related to dust dynamics not included in our models is the potential for dust traps. High-resolution ALMA continuum observations show that many discs are structured \citep{Andrews2018}. These structures are likely due to the presence of dust traps, which may change the disc composition as a result of trapping molecular ices in the outer disc \citep[e.g.][]{Kama2015, Sturm2022, Kalyaan2023}. Due to the ordering of ice lines, this could lead to low oxygen abundances and high C/O ratios in the inner disc, likely making it appear as if the planets formed further from their star than they otherwise did.

\section{What might the BOWIE-ALIGN sample tell us about planet formation?}\label{sec:insight}

If the BOWIE-ALIGN survey of hot Jupiters shows differences between the disc migration (aligned) and high-eccentricity migration (misaligned) planets, as predicted by the simplified or baseline models, then this sample likely tells us something about how these planets formed or what conditions they formed in. 
The model distribution in Fig.~\ref{fig:C_set} highlights that the current uncertainties of the physics and chemistry of disc and planet allow a wide range of formation models, that only observations can narrow down to the actual process of planet formation.
Here, we summarize what the main takeaway points would be, accounting for the complications discussed in \autoref{sec:assumption}.
Previous observations of hot gas giants show super-solar metallicities \citep[e.g.,][]{Alderson2023,Ahrer2023,Bean2023,Feinstein2023,Rustamkulov2023,Xue2024}. This hints at a significant accretion of solids such as planetesimals, but for moderately super-solar metallicity (a few times solar), the models are degenerate because super-solar metallicity gas produced by efficient pebble accretion and drift is also possible. 
The pebbles in the models for atmospheric assembly mainly enrich the accreted gas when reaching the inner disc rather than being directly accreted as the final planet mass is several times larger than the pebble isolation mass.
Together with the C/O ratios, the metallicity may lead to three scenarios with three distinct conditions.

\begin{itemize}
    \item High planetesimal accretion: If we measure sub-solar C/O ratios that show a split in the C/O ratio where the aligned planets have a lower C/O, the formation of such planets would have required solid accretion enriching the upper atmosphere late in the evolution. Such accretion happens through planetesimals.
    \item High pebble flux: If we find higher C/O ratios (around solar or above) in the observational sample with no clear distinction between aligned and misaligned planets (assuming the observations reach sufficient precision to separate C/O ratios with differences at the 0.1 to 0.2 level), then this may point to planet formation in a disc with metal-rich gases and efficient radial drift of solids.
    \item Poorly mixed interiors:
    If the metallicity is only a few times super-solar and C/O ratios are $>0.5$, then the two migration scenarios can produce planets with different C/O ratios (\autoref{fig:fin_set}) if the interior mixing of the envelope is inefficient and gas accretion dominates: aligned planets have C/O below values of $<1$ and misaligned planets $>1$.
    \item Silicate condensation: If we find that planets tend to have higher C/O ratios ($>0.6$, for example), with the misaligned planets being more carbon depleted, the oxygen bound in silicates might not have been released into the atmosphere. In this case, the atmospheric composition will have been dominated by accreted gas and volatile species. Moderately volatile species such as sulphur might then hold the key to probing the formation history of these planets \citep[e.g.][]{Turrini2021}. 
\end{itemize}

\begin{figure}
{\includegraphics[width=\columnwidth]{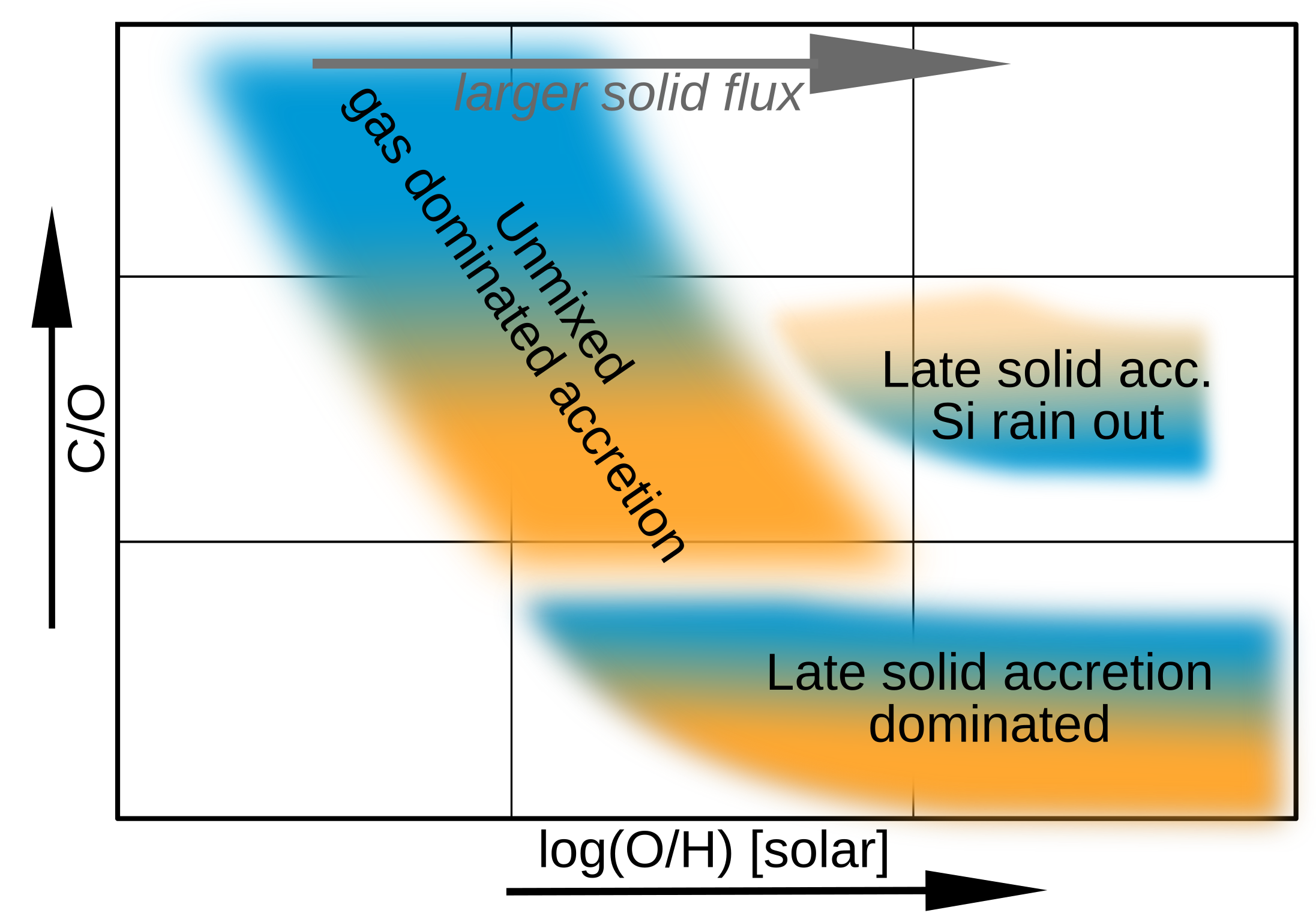}}
{\caption{Simple sketch to highlight the regions, that could be used to understand formation processes if the observed sample falls completely in one of the regions. Aligned planets are represented with orange, misaligned planets with blue. Uncoloured regions are degenerate model solutions and do not allow an interpretation. The specific values of the trends can shift due to model assumptions while maintaining the trends.}
\label{fig:sketch}}
\end{figure}

We visually summarise these possibilities in the sketch in Fig.~\ref{fig:sketch}. It highlights how the population of different migration paths in the BOWIE-ALIGN programme can help us to learn more about planet formation.

\section{Conclusions}\label{sec:conclusions}

In the BOWIE-ALIGN programme \citep{2024survey} we will use JWST to compare the atmospheric compositions of four hot Jupiters that likely underwent disc migration (inferred from their aligned orbits around F stars) with four that likely migrated after the dispersal of the disc via high-eccenticity migration (inferred from their misaligned orbits). Here, we investigate how the different formation histories of the two populations can change the resulting metallicity and C/O ratio using a simple model of planet formation with a wide range of disc and accretion parameters.
Many aspects of planet evolution are still uncertain, including conditions in the disc, such as the temperature, composition,  disc masses, and dust properties, and the importance of different planet formation processes including planet migration and pebble accretion. Here we show how the predicted compositions of the aligned and misaligned planets vary with different disc physics assumptions.

Our model reproduces the basic findings of previous works. For example, hot Jupiters that arrived at their final locations by migrating through the disc likely acquire bulk compositions that have lower C/O ratios than the high-eccentriciy migration sample \citep[e.g.][]{Madhusudhan2014b}. This is a result of the lower C/O ratio of solids in the inner disc. Similarly, we recover a sample of planets with high C/O ratio and moderately super-solar metallicity when including the efficient pebble drift associated with pebble accretion models \citep[e.h.][]{Booth2017, Schneider2021}.

However, we identify new possibilities which have not been considered before:

\begin{itemize}
    \item High metallicity ($[\mathrm{O/H}] > 10\times$ solar) likely requires the accretion of a significant amount of solids through planetesimals. If these planetesimals are completely disrupted and contribute to the atmospheric composition then high metallicity planets will have sub-solar C/O ratios, typically with C/O$<0.3$. Without significant enrichment by planetesimals, atmospheric metallicities are below $5\times$ solar while C/O is typically super-solar.
    \item At high metallicities ($\gtrsim 5 \times$ solar), the aligned (disc-migrated) and misaligned (high-eccentricity migrated) planets have different C/O ratios, with the aligned planets being consistently more carbon depleted than the misaligned planets.
    \item If the planet's interior mixes weakly then differences between the aligned and misaligned planets widen, particularly at high [O/H]. This arises because the atmospheric composition traces only the 
    short distance of the final growth phase when mixing is weak. 
    \item The fate of silicates significantly affects the planets' C/O ratio. Due to the high condensation temperature of silicates, they may condense and rain out the planets atmosphere during formation, lowering the observed oxygen abundance due to the oxygen contained in silicates. Silicate condensation can cause the C/O ratio of the planets to be super-solar and even flip the sign of the difference between the two samples, making the aligned hot Jupiters more carbon rich. This arises because the carbon grains are the most important contribution to the solid abundance in the inner discs after silicates. The amount of carbon in refractory carbon grains, and their ability to survive in the inner disc, needs to be better constrained to properly understand their impact on planet compositions. 
\end{itemize}

These challenges mean it is difficult to predict a priori how the aligned and misaligned samples of planets will differ. Nevertheless, atmospheric observations of a population of planets with known differences in their history offers the best opportunity for understanding which of these factors are significant. 

\section*{Acknowledgements}
RAB and JEO thank the Royal Society for the support in the form of University Research Fellowships, ABTP received support from their Enhancement Awards. This project has received funding from the European Research Council (ERC) under the European Union’s Horizon 2020 Framework Programme (grant agreement no. 853022, PEVAP). JK acknowledges financial support from Imperial College London through an Imperial College Research Fellowship grant. N.J.M., D.E.S and M.Z. acknowledge support from a UKRI Future Leaders Fellowship [Grant MR/T040866/1], a Science and Technology Facilities Funding Council Nucleus Award [Grant ST/T000082/1], and the Leverhulme Trust through a research project grant [RPG-2020-82]. PJW acknowledges support from the UK Science and Technology Facilities Council (STFC) through consolidated grants ST/T000406/1 and ST/X001121/1. V.P. acknowledges support from the UKRI Future Leaders Fellowship grant MR/S035214/1 and UKRI Science and Technology Facilities Council (STFC) through the consolidated grant ST/X001121/1.

\section*{Data Availability}
The data underlying this article will be shared on reasonable request to the corresponding author.
The code to reproduce the models can be found here: \url{https://github.com/miosta/drift_composition/}

\bibliography{planet_composition}
\bibliographystyle{mnras}

\appendix

\section{Planet model}
\label{sec:planet}

\subsection{Disc model}\label{sec:disc}

We model the disc as a steady-state, dusty, viscous accretion disc with molecules that can condense and sublimate from the grain surfaces. The model's key physical and compositional parameters are given in \autoref{tab:models} and \autoref{tab:chem_abund}.

In steady-state the gas accretion rate, $\dot{M}_{\rm g}$, is constant and given by
\begin{equation}
\dot{M}_{\rm g} = 3\upi \nu \Sigma.
\end{equation}
We use a Shakura-Sunyaev viscosity model, $\nu = \alpha c_s H$ with $H = c_s / \Omega_{\rm K}$ where $\Omega_{\rm K}$ is the Keplerian angular velocity. From this, we compute the gas surface density, $\Sigma$, and the gas radial velocity, $v_{\rm R, g} = - 1.5 \nu/R$. The mid-plane temperature profile is specified to be $T = T_0 (R/R_0)^{-0.5}$, where $R_0 = 1\,{\rm au}$. 

Dust is assumed to drift radially towards the star due to the sub-Keplerian rotation of the pressure-supported gas. The dust radial velocity is taken to be
\begin{equation}
v_{\rm R, d} = \frac{v_{\rm R, g} - \eta v_{\rm K} St}{1 + St^2},
\end{equation}
where $\eta = -(R/\rho)dP/dR$ and $v_{\rm K}$ is the Keplerian velocity \citep{Takeuchi2002}. Here, $P$ and $\rho$ are the mid-plane pressure and density, computed assuming that the vertical structure of the disc is isothermal and a Gaussian. The Stokes number, $St$, is assumed to be constant. Where needed, the sizes of the grains, $a$, are computed assuming an Epstein drag law \citep{Epstein1924} evaluated at the disc mid-plane via $St = 0.5\upi\rho_s a/\Sigma$, where $\rho_s = 1\,{\rm g\,cm^{-3}}$ is the internal density of the grains. 

The surface densities of dust and molecules in both gas and solid phases are computed by solving advection-diffusion-reaction equations. For the gas-phase species with surface densities $\Sigma_{\mathrm{g},i}$,
\begin{align}
\prt{\Sigma_{\mathrm{g},i}}{t} +& \frac{1}{R} \frac{\partial}{\partial R}\left(R\, v_{\rm R, g} \Sigma_{\mathrm{g},i}\right) = \nonumber \\& \frac{1}{R}\frac{\partial}{\partial R}\left [R\, \frac{\nu \Sigma}{Sc} \frac{\partial}{\partial R}\left(\frac{\Sigma_{\mathrm{g}i}}{\Sigma}\right)\right] + \dot{\Sigma}_{{\rm S},i}-\dot{\Sigma}_{{\rm C},i}, \label{eq:adv-diff-react-gas}
\end{align}
and 
\begin{align}
\prt{\Sigma_{\mathrm{d},i}}{t} +& \frac{1}{R} \frac{\partial}{\partial R}\left(R\, v_{\rm R, d} \Sigma_{\mathrm{d},i}\right) = \nonumber \\ &\frac{1}{R}\frac{\partial}{\partial R}\left [R\, \frac{\nu \Sigma}{Sc} \frac{\partial}{\partial R}\left(\frac{\Sigma_ {\mathrm{d},i}}{\Sigma}\right)\right] + \dot{\Sigma}_{{\rm C},i}-\dot{\Sigma}_{{\rm S},i},
\label{eq:adv-diff-react-dust}
\end{align}
for the dust and ices with surface densities $\Sigma_{\mathrm{d},i}$. The sublimation and condensation rates of the ices are denoted  $\dot{\Sigma}_{\rm{S},i}$ and $\dot{\Sigma}_{\rm{C},i}$, respectively (which are assumed to be zero for the silicate dust grains). We solve these equations under the assumption of steady-state (i.e. $\partial\Sigma_i/\partial t=0$). Here $\Sigma_i$ refers to the surface density of the species in question, the Schmidt number, $Sc$, is the ratio of the viscosity and diffusion coefficients, and  $Sc=1$ is assumed. The final terms in the equations encode the sublimation and condensation terms;  for the dust density we set these to zero. We solve these equations using a finite-volume method on a grid covering the domain 0.001--100~au with 512 logarithmically-spaced radial cells.

The condensation rate, $\dot{\Sigma}_{{\rm C},i}$, is determined by the collision rate of gas-phase molecules with the grain surfaces. We compute this assuming that the dust is settled to the mid-plane. The vertically integrated-condensation rate is then given by
\begin{equation}
\dot{\Sigma}_{{\rm C},i} = \frac{\Sigma_{\rm d} \langle \upi a^2 \rangle}{\langle m_{\rm d}\rangle} \frac{\Sigma_{{\rm g}, i}}{\sqrt{2\upi}H} \sqrt{\frac{8k_B T}{\upi m_{i}}} P_{\rm stick}.
\end{equation}
The first term is the total integrated surface area of the dust grains computed from the dust surface density, the average mass of a grain ${\langle m_{\rm d}\rangle}$ and the average grain area $\langle \upi a^2 \rangle$. These are computed assuming an size distribution according to \cite{MRN}(MRN) ($n(a) \propto a^{-3.5}$) with the maximum grain size determined by the Stokes number. The second term gives the mid-plane mass density of the molecule $i$ in the gas phase (assuming efficient vertical mixing), while the third term is the mean thermal velocity of the molecules ($m_i$ is the mass of the molecule being considered), and the last term is the probability that the molecule sticks to the grain surface, which we take to be 1.

The sublimation rate, $\dot{\Sigma}_{{\rm S},i}$, is modelled using the standard kinetic approach \citep[see, e.g.,][]{2001Fraser,  Bisschop2006, Cuppen2017}. Assuming that the temperature does not vary with height, the vertically integrated sublimation rate is
\begin{equation}
\dot{\Sigma}_{{\rm S},i} = \frac{\Sigma_{\rm d} \langle 4\upi a^2 \rangle}{\langle m_{\rm d}\rangle} N_{\rm bind} \, m_i f_{\rm C, i} \, \nu_i \exp\left(- \frac{E_{\rm sub,i}}{T}\right),
\end{equation}
where $N_{\rm bind} = 10^{15}\,{\rm cm}^{-2}$ is the number of binding sites per unit area. $\nu_i$, is the attempt frequency and $E_{\rm sub}$ is the binding energy (in Kelvin), which are given in \autoref{tab:chem_abund}. The covering fraction, $f_{{\rm C},i}$, is the fraction of binding sites covered by the ice. For simplicity, we take this to be
\begin{equation}
f_{\rm C, i} m_i = \frac{\Sigma_{{\rm d},i}}{\Sigma_{{\rm d}, i} + \frac{\Sigma_{\rm d} \langle 4\upi a^2 \rangle}{\langle m_{\rm d}\rangle} N_{\rm bind} m_i},
\end{equation}
where the second term in the denominator is the number of binding sites per unit area in the disc. This smoothly interpolates between the first-order desorption rate when the ice is less than a single layer thick and the zeroth-order desorption rate appropriate for multiple layers of ice.

The final process we include is carbon grain destruction. We model this following \citet{2017Gail}, in which we assume that the carbon grains thermally decompose at a rate $$\dot{\Sigma}_{\rm C-grain} = -\Sigma_{\rm C-grain} \nu_{\rm C-grain} \exp(-E_{\rm sub,C-grain}/T),$$ where the constants are given in \autoref{tab:chem_abund}. The decomposed carbon grains are added to the gas and are not allowed to recondense because the thermal decomposition is irreversible. Note that \citet{2017Gail} considers three different types of refractory organics that decompose at different rates. Here, we use only their `volatile organics' for simplicity.

Since we compute steady-state solutions, the accretion rate of the gas, $\dot{M}_{\rm g}$, the dust, $\dot{M}_{\rm d}$,  and for each molecule must be specified at the outer boundary. We take $\dot{M}_{\rm g}$ and $\dot{M}_{\rm d}$ to be free parameters, and determine $\dot{M}_{i}$ for each molecule from its abundance. For molecules that are in the gas phase at the outer boundary (only H$_2$ and the noble gases), we set $\dot{M}_{i} = X_i \dot{M}_{\rm g}$, where $X_i$ is the mass-fraction of species $i$. For species that are in ice form we use $\dot{M}_{i} = X_i \dot{M}_{\rm d} / X_{\rm d}$, where $X_{\rm d}$ is the dust-to-gas ratio in the disc. This form is based on the assumption that as the accretion rate of dust increases or decreases, the accretion rate of ice should increase or decrease accordingly. As a result, if $\dot{M}_{\rm d} / \dot{M}_{\rm g}$ differs from the dust-to-gas ratio, $X_{\rm d}$, then the metallicity of the disc will change accordingly. $\dot{M}_{\rm d} / \dot{M}_{\rm g}$ may plausibly take higher or lower values than $X_{\rm d}$: efficient radial drift increases the ratio, but as a result the dust mass in the disc can decrease rapidly, leading to low values of $\dot{M}_{\rm d} / \dot{M}_{\rm g}$ at late times \citep[e.g.][]{Booth2020}.

\subsection{Planet growth}

In our model, the planet will grow from solids and gases present in the disc.
For the accretion of solids we consider two pathways: pebble accretion as described in \cite{2015Bitsch} and planetesimal accretion simplified from the results of \cite{2013Fortier}.
As we are mainly interested in the final assembly of a gas giant atmosphere, we start the simulation with super-Earth planets with a fraction of the mass in the atmosphere. 

\subsubsection{Pebble Accretion}
Pebble accretion is split into two regimes; they can be accreted from a cross-section that is either i) fully embedded in the pebble layer in the disc or ii) extends above and below the thickness of the pebble layer.  
The transition between the two regimes is given by \citet{2015Bitsch}:
\begin{equation}
    r_{\mathrm{Hill},c} = H_\mathrm{peb} \sqrt{\frac{8}{\upi}} (\frac{\St}{0.1})^{-1/3},
\end{equation}
where $r_\mathrm{hill}=r_p (1/3\cdot m_p/M_\star)^{1/3}$ is the Hill Radius,  $m_p$ is the current planet mass, and $r_p$ its semi-major axis. The pebble scale height $H_\mathrm{peb}$ is be calculated assuming equilibrium between turbulence and settling, 
$H_\mathrm{peb} = H /  \sqrt{1 + \St/\alpha}$.

In case ii) pebble accretion occurs in a 2D fashion, for which the accretion rate is given by
\begin{equation}
    \dot{m}_\mathrm{2d,peb} = 2 (\frac{\St}{0.1})^{2/3} r_\mathrm{Hill}^2 \Omega_\mathrm{K} \Sigma_\mathrm{d}.
\end{equation}
In regime i) the accretion rate is reduced as only part of the pebble layer can be accreted. This is usually referred to as the 3D regime and the pebble accretion rate is given by
\begin{equation}
    \dot{m}_\mathrm{3d,peb} = \dot{m}_\mathrm{2d} \frac{r_\mathrm{Hill}}{H_\mathrm{peb}} \sqrt{\frac{\upi}{8}}(\frac{St}{0.1})^{1/3}.
\end{equation}
The 3D regime is only important for low-mass planets but can become important at large distances in a flared disc where the disc's aspect ratio exceeds 5\% significantly.

Growth by pebble accretion stops once the planets reach the pebble-isolation mass, $M_{iso} = 20 M_\oplus (h/0.05)^3$, where $h=H/R$ is the disc aspect ratio.

\subsubsection{Planetesimal Accretion}

We also consider planet growth due to the accretion of planetesimals using a prescription based on the model presented in \citet{2013Fortier}. The accretion rate is parameterised by the rate at which planetesimals enter the planet's Hill sphere due to Keplerian shear, multiplied by a capture probability, $P_\mathrm{col}$. Denoting the surface density of planetesimals in terms of the local pebble density, $\Sigma_{\rm pl} = f_{\rm pl} \Sigma_{\rm f}$, the accretion rate via this mechanism is simply given by:
\begin{equation}
    \dot{m}_\mathrm{pls} = P_\mathrm{col} f_\mathrm{pl} \Sigma_\mathrm{d} \rh^2 \frac{\Omega_\mathrm{K}}{2\pi}
\end{equation}

\begin{figure}
{\includegraphics[width=\columnwidth]{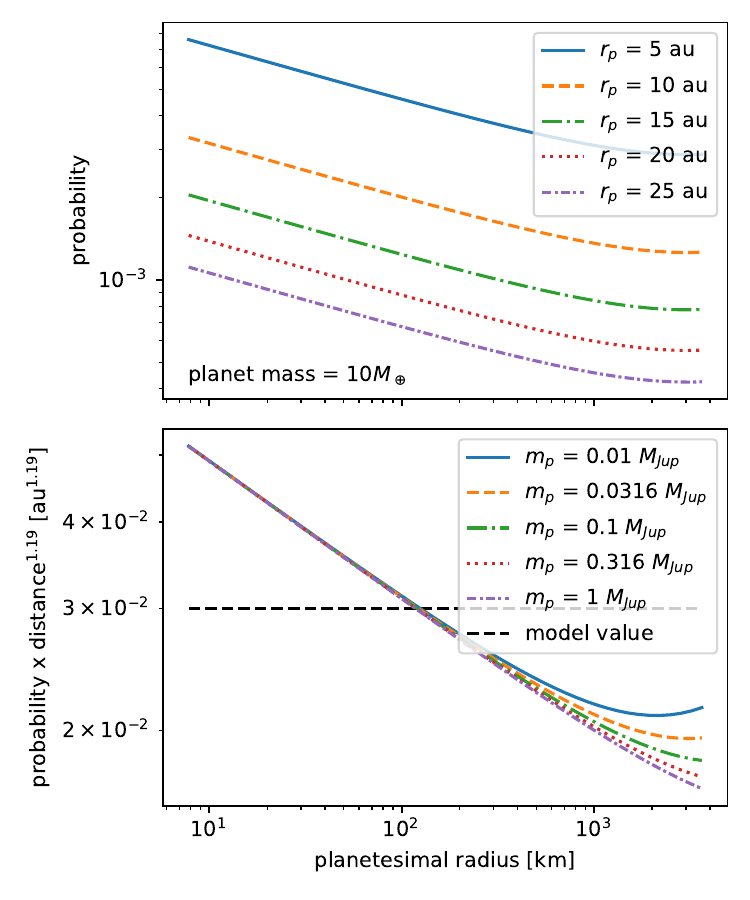}}
{\caption{The collision probability of planetestimals as a function of planetesimal mass. Top panel: the dependence of probability on radius / separation for a fixed mass of 10\,M$_{\oplus}$. Bottom panel: The dependence of probability on planet mass for a fixed distance.}\label{fig:plansi}}
\end{figure}

While the $P_{\rm col}$ is a function of the planetesimal size and the resulting distribution of orbit parameters, for planetesimals with sizes ranging from a kilometre in radius to a $0.01 M_\oplus$ object, the resulting probabilities change less than by a factor 3 as shown in Fig.~\ref{fig:plansi}. Since we consider a range of planetesimal-to-dust ratio $f_\mathrm{pl}$ in the range from $10^{-5} - 1$, the variation in $P_{\rm col}$ that arises from considering different planetesimal size distributions is insignificant.
Therefore for our model, we will consider a single probability of an average size planetesimal with 60~km of $\sim 10^{-6}~M_\oplus$.

The collision probability also depends on the radial distance due to its relation to the Hill sphere, the settling of the eccentric and inclined motion of the planetesimals at closer-in orbits in the disc and the temperature and Hill radius dependent extent of the planet \citep{2013Fortier} as shown in the top panel in Fig.~\ref{fig:plansi}. The resulting radial dependence of the collision probability can be approximated with a simple radially dependent power-law function.
\begin{equation}
    P_\mathrm{col} = 3\times 10^{-2}(1\mathrm{AU}/r_\mathrm{p})^{1.19}.
\end{equation}
We limit it further to $P_\mathrm{col}\leq 3\times10^{-2}$. This limits the accretion of planetesimals inside the final au, where the power-law function would otherwise lead to an unreasonable increase in planetesimal accretion rate.

\subsubsection{Gas Accretion}
The main part of accretion will consist of gas accretion. Westart with rocky super-Earths in which the core mass exceeds the atmosphere mass, so we consider both the hydrostatic growth phase and runaway gas accretion. We start in this regime to eliminate planets that do not reach runaway gas accretion, as their feeding zone might be too small or feeding too slow.
When $m_\mathrm{core} < m_\mathrm{gas}$ the gas accretion is in the hydrostatic regime for which we use the gas accretion rates based on the Kelvin-Helmholtz time-scale from \citet{Piso2014}:
\begin{align}
    \dot{m}_\mathrm{gas} = &0.00175 f^{-2}\frac{\kappa}{1~\mathrm{cm^2/g}} \sqrt{\frac{81~\mathrm{K}}{T}} \nonumber \\
    &(\frac{\rho_\mathrm{core}}{5.5~\mathrm{g/cm^3}})^{-1/6} (\frac{m_\mathrm{core}}{M_\oplus})^{11/3} \frac{0.1 M_\oplus}{m_\mathrm{gas}} \frac{M_\oplus}{\mathrm{Myr}}
\end{align}

When the gas mass exceeds the core mass, the planet enters runaway accretion. To model runaway gas accretion, we use the results of hydrodynamic simulations from \citet{2010Machida}. \citet{2010Machida} conducted high-resolution shearing-box simulations of runaway gas accretion, for which the local gas surface density was prescribed. Since giant planets can carve deep gaps in the disc, we need to account for the depleted gas surface density when calculating the accretion rate. Since we do not consider the impact of the planet on the disc models we account for gap opening during gas accretion by scaling the gas surface density near the planet according to the prescription in \citet{2018Kanagawa}:
\begin{align}
    \Sigma_\mathrm{gap} =& \Sigma_\mathrm{g} (1+0.04K)^{-1}\nonumber \\
                & \mathrm{with}\quad K = (\frac{m_p}{M_\star})^2 \frac{1}{h^5\alpha}
\end{align}
Using $\Sigma_\mathrm{gap}$ as the surface density in the parameterisation of \citet{2010Machida}, we arrive at the following expression for the gas accretion rate in the two runaway regime:
\begin{align}
    \dot{m}_\mathrm{gas,3d} &= 0.83 \Omega_\mathrm{K} \Sigma_\mathrm{gap} H^2 (\frac{\rh}{H})^{9/2} \\
    \dot{m}_\mathrm{gas,2d} &= 0.14 \Omega_\mathrm{K} \Sigma_\mathrm{gap} H^2,
\end{align}
where the minimum of these is used.
The resulting gas accretion rates are comparable to the gas accretion found in the study by \cite{2023Li}. We also limit the gas accretion to a fraction of the mass flux through the disc.
The composition of the accreted gas follows the gas composition in the disc at position of the planet.

\subsection{Planet migration}

As we are mainly interested in the late stages of the planet's evolution, when the bulk gas is accreted, we base our planet migration prescription on the numerical simulations of \citet{2015Durmann}, who focussed on gap opening planets close to the classical Type II regime. In the classical Type II regime, the planet is expected to migrate at a velocity close to the viscous velocity, $v_{\rm R, g}$. We scale the migration speed to this using a mass factor, $f_{\rm mp}(m_{\rm p})$, and a density factor, $f_{\Sigma}(\dot{M}_{\rm g})$ to make the results of \citet{2015Durmann}. The resulting prescription is:
\begin{align}
    \dot{a} =& v_\mathrm{R, g} f_\mathrm{mp} f_\Sigma \nonumber \\
            &\mathrm{with} \quad f_\mathrm{mp}= \mathrm{min}(0.09 (\frac{m_\mathrm{p}}{M_\mathrm{Jup}})^{-0.4}; 5) \nonumber \\
            &\mathrm{and} \quad f_\Sigma= \mathrm{min}(4 (\frac{\Sigma_\mathrm{g}(r)}{\Sigma_\mathrm{std}(r)})^{0.6}; 5)
\end{align}

The $\Sigma_\mathrm{std}$ is derived from a mass accretion rate of $10^{-7}~M_\odot/\mathrm{yr}$ as $\Sigma_\mathrm{std}(r) = 10^{-7}~M_\odot/\mathrm{yr} / (3\pi\nu)$.
In addition, we limit the migration to a minimum of $10\%~v_{\rm R, g}$. Thereby, the planet will initially migrate much quicker than its final speed as a giant. While this is a crude approach to the migration, especially for low masses, it is sufficient to filter out initial planets and locations that do not allow the planet to enter into runaway accretion due to the initial migration during the onset of gas accretion.

\subsection{Time integration}

The planets are integrated using a first order explicit integration. To ensure migration and growth are well-resolved, the timestep is adaptive to the migration velocity 
$v_\text{mig}$ using the disc cell size $dR$ and the mass accretion $\dot{m}$ with a change of the current mass $m$ by $\leq5\%$ by finding:
\begin{equation}
     dt = \text{min}(0.4 \frac{dx}{v_\text{mig}}, 0.05 \frac{m} {\dot{m}})
\end{equation}
The maximal allowed timestep is $10^4$~yrs.

For each planet, we follow their growth until one of three conditions are met:
\begin{itemize}
    \item The planet reaches the final orbit, randomly chosen in the range 0.01--0.1~au.
    \item The planet mass reaches $2~M_\text{Jup}$.
    \item The simulation reached $10^7$~yrs.
\end{itemize}

\begin{table}
    \centering
    \begin{tabular}{l|c|c|c|r}
    Molecule & Abundance [$X/H$] & $\nu_{i}~[s^{-1}]$ & $E_\mathrm{sub}~[K]$ & Ref.\\
    \hline
    H$_2$O  & $1.963 \times 10^{-4}$ &  $4.0\times 10^{13}$ & 5800   & (1) \\
    CO$_2$  & $4.908 \times 10^{-5}$ &  $1.0\times 10^{13}$ & 2700   & (2) \\
    CO   &$9.815 \times 10^{-5}$ &  $7.0\times 10^{11}$ & 1180   & (3) \\
    CH$_4$  &$4.410 \times 10^{-5}$ &  $1.1 \times 10^{12}\,^*$ & 1250   & (4) \\
    C$_2$H$_6$ &$(2.205 \times 10^{-5})\,^+$ &  $6.0\times 10^{16}$ & 2500   & (5) \\
    CH$_3$OH &  & $1.6 \times 10^{12}\, ^*$ & 4930   & (4)  \\
    N$_2$   &$3.658 \times 10^{-5}$ & $8.0\times 10^{11}$ & 1050   & (3) \\
    NH$_3$  &$8.129 \times 10^{-6}$ &  $1.0\times 10^{13}$ & 3800   & (6) \\
    He   &$9.539\times 10^{-2}$ &  $6.5\times 10^{11}\,^*$ & 100    & (4) \\
    Ar   &$3.020\times 10^{-6}$ &  $6.0\times 10^{11}$ & 870    & (7) \\
    Kr   &$2.165\times 10^{-9}$ &  $1.2\times 10^{14}$ & 1380   & (7) \\
    Xe   &$2.044\times 10^{-10}$ & $4.6\times 10^{14}$ & 1970   & (7)\\
    C-grain &$1.323\times 10^{-4}$ &  $4.0\times 10^{13}$ & 19050  & (8)  \\
    Silicates & $2453 \times 10^{-5}$  & NA & NA & \\
    \end{tabular}
    \caption{Table of the considered species. The abundances in the fiducial disc model are given in the second column; the carbon and oxygen abundances are determined as in \protect\citet{Oberg2019}, using the solar reference \protect\citep{Asplund2009}. The desorption rates $\nu_{i}$ and sublimation energies $E_\mathrm{sub}$ use the following references: \protect\cite{2001Fraser}$^1$, \protect\cite{1990Sandford}$^2$,\protect\cite{2016Fayolle}$^3$,\protect\cite{2017Penteado}$^4$,\protect\cite{2019Behmard}$^5$, \protect\cite{2015Suhasaria}$^6$, \protect\cite{Smith2016}$^7$,\protect\cite{2017Gail}$^8$.
    The asterisks mark theoretical estimates following \protect\citet{Tielens1987} rather than experiment findings. The ethane abundance (+) is zero except for models that use ethane instead of methane.}
    \label{tab:chem_abund}
\end{table}

\section{Effects of free parameters} \label{sec:parameters}

\begin{figure*}
{\includegraphics[width=2\columnwidth]{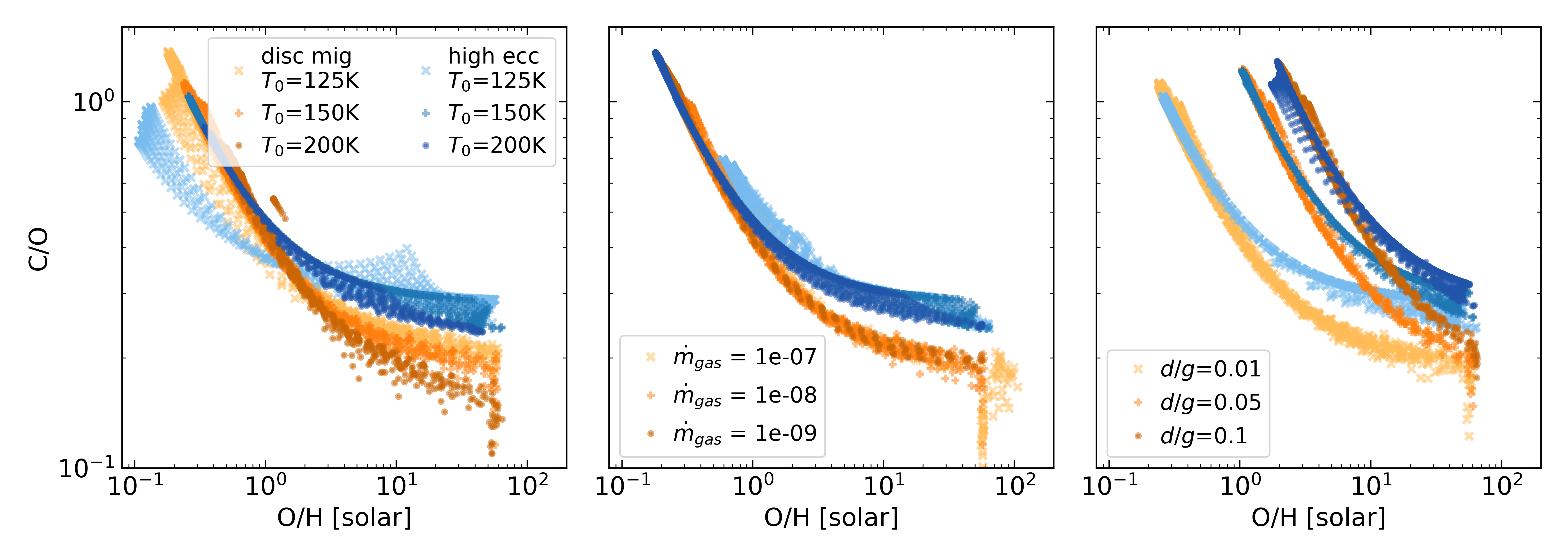}}
{\caption{Sample for the effects of temperature, disc mass and dust-to-gas flux. Blue colour indicates high eccentricity migration, orange colour indicates disc migration.
The left most panel probes the fiducial setup for 3 different temperature profiles from bright to dark 125K, 150K and 200K at 1au. The middle panel looks at 3 different mass accretion rates for bright to dark: $[10^{-7}, 10^{-8}, 10^{-9}]$. The right panels models dust-to-gas flux ratios of [1\%, 5\%, 10\%].}\label{fig:samples_parameters}}
\end{figure*}

Three of the parameters varied in our study are temperature $T_0$, disc mass $M_\text{g}$ and dust-to-gas flux ratio $f_\text{d/g}$.
These have minor effects on the sample of planets of different initial radii and planetesimal fractions, as shown in Fig.~\ref{fig:samples_parameters}.
We examine each parameter's effects on planet evolution one by one to understand the shift they produce.

\subsection{Temperature}\label{sec:Temp}

\begin{figure*}
{\includegraphics[width=2\columnwidth]{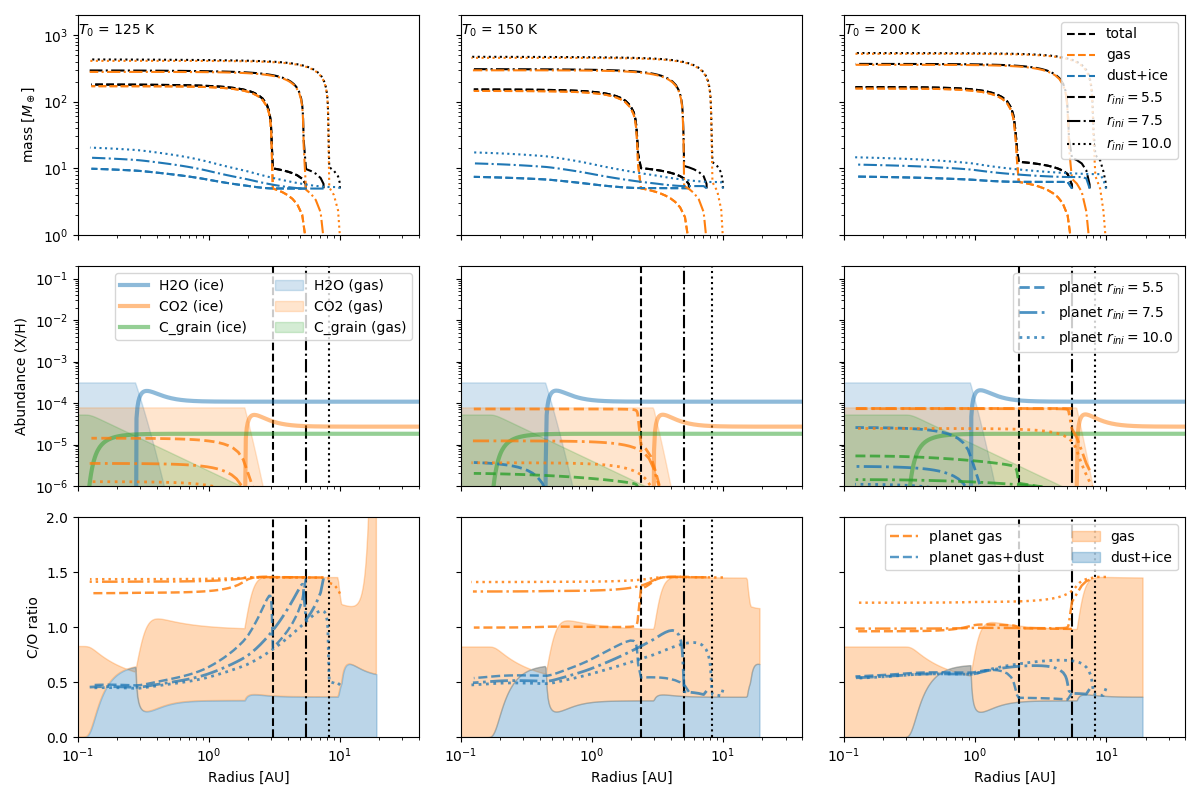}}
{\caption{Planet and composition evolution for 3 different temperatures. The columns show models for left to right for 125K, 150K and 200K. The top row shows the planet growth of planets with starting at 5.5, 7.5 and 10~au indicated by the line style. Black lines indicate the total growth, blue indicates solid accretion and orange lines indicate gas accretion. The middle row shows the gas abundance of carbon refractories (green), water (blue) and CO$_2$ (orange) in surfaces and the solid abundance in thick solid lines. The gas abundance of molecules accreted by the planets are indicated by the lines style corresponding to the top row. The black lines indicate the runaway location or the planets. The bottom panel shows the disc C/O ratio in orange surfaces for gas and blue surfaces for the solid components. The orange lines indicate the C/O ratios of the accreted gas and the blue lines the C/O ratio of the solid and gas accretion.}\label{fig:ex1}}
\end{figure*}

We explore the effects of the disc temperature profile by changing $T_0$, the temperature at 1~au, but keeping the scaling with radius fixed. Changing the temperature affects the positions of the ice lines, but not the composition between them (middle panel Fig.~\ref{fig:ex1}). As a result, there is a shift in the planets' composition in Fig.~\ref{fig:samples_parameters} in the C/O (towards low C/O at high metallicity and vice-versa), but the overall trends remain the same.

One of the most important phases is the onset of the runaway gas accretion, in which a planet gains a large fraction of its final mass and atmosphere in gas quickly while migrating only a short distance. 
The location of runaway accretion thereby has a large influence on the planet's final composition. 
The balance of growth and migration sets the location of the onset of the runaway accretion.

One of the most significant changes in abundance happens around the icelines of CO and CH$_4$ in our model. If runaway accretion starts outside these ice lines the planets accrete gas that has carbon and oxygen abundances depleted by an order of magnitude. For the coldest disk models ($T_0 = 125$~K), in which a significant number of planets begin runaway gas accretion outside these icelines, this results in a population of very low metallicity planets (down to $0.1\times$ solar) at super-solar C/O ratios. At higher temperatures, the planets typically undergo runaway gas accretion inside these icelines, resulting in a smaller range of metalliticities at a given C/O ratio. There does, however, remain a small but systematic difference between the compositions of the planets formed in the fiducial and warmer discs.

The temperature also affects a number of processes in the formation and migration of the planets. For example, the temperature sets the gas scale height of the disc, $H$, which determines the pebble isolation mass. The large pebble isolation mass at the higher temperatures (Fig.~\ref{fig:ex1} right) allows for faster growth and an earlier onset of runaway gas accretion.
The higher pebble content buffers the planet against the change in C/O during the early phases of runaway gas accretion, resulting in a smaller range of metalicities and C/O ratios for the planets forming in warmer discs. 
Since planet migration is limited by the viscosity, which increases with temperature, warmer discs result in faster migration.
Hence, in the coldest case (Fig.~\ref{fig:ex1} right), the planets migrate more slowly and the onset of run-away accretion is about 0.5~au further (given the same initial location).

In the late stage of evolution, gas accretion is limited due to the hill sphere shrinking with migration, and the deep planetary gap. Meanwhile the high orbital speed leads to more planetesimal collisions to enrich the planets. Hence, the total gas+solid C/O of the planets declines until the water iceline is crossed (see the bottom panel Fig.~\ref{fig:ex1}).

The combination of these effects means that increasing the temperature of the disc results in a slight reduction in C/O of all the planets. It also remains true that the high-e migrated planets at high metallicities at lower C/O ratios than the disc migrated planets when considering the range of temperatures.

\subsection{Disc mass and dust-to-gas ratio}

\begin{figure}
{\includegraphics[width=\columnwidth]{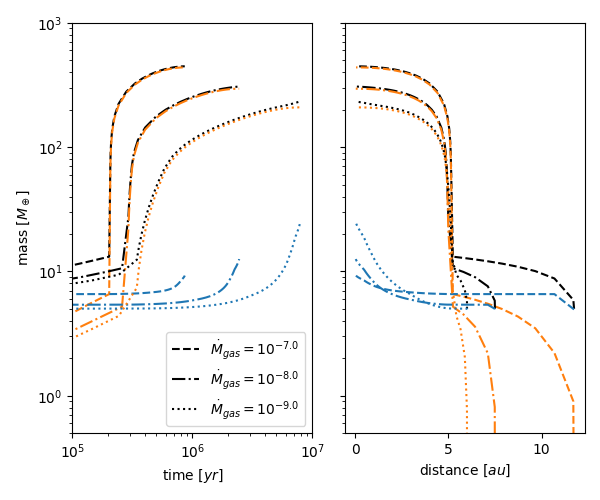}}
{\caption{Evolution track of planets with different disc masses. Black represents the total growth, orange the gas accretion only and blue the solid accretion only. The different line styles correspond to discs with different mass accretion rates of $10^{-7}$\,$\mathrm{M_\odot/yr}$ (dashed lines), $10^{-8}$\,$\mathrm{M_\odot/yr}$ (dashed-dotted lines), and $10^{-9}$\,$\mathrm{M_\odot/yr}$ (dotted lines).  All planets started such that their runaway accretion sets in at the same time.}\label{fig:example_mdot}}
\end{figure}

The disc mass affects the planet migration speed and accretion rate of solids and gas, which are all processes scaling with the surface density in the disc. However, the faster migration speed is more significant resulting in lower final planet metallicities in discs with a higher mass (or accretion rate), as shown in Fig.~\ref{fig:example_mdot}. Nevertheless, the impact on the \emph{population} of the planets is quite small as the final effect is degenerate with changing the planetesimal mass fraction (Fig.~\ref{fig:samples_parameters}). Some differences do arise, however, because planets with the same final location undergo runaway gas accretion further out in the disc. Therefore higher disc masses / accretion rates produce more planets with higher C/O ratios than discs with lower masses.

Increasing the dust-to-gas flux ratio shifts the overall distribution towards more metal-rich planets, as there are more solids to accrete and the gas phase abundances increase due to more ice evaporating at the icelines. The right panel in Fig.~\ref{fig:samples_parameters} shows the shift.
Increased dust-to-gas flux ratios naturally arise due to efficient radial drift \citep[e.g.][]{Booth2020}. As a result, the lowest metallicity planets are likely unrealistic since either a significant concentration of planetesimals or large pebble fluxes are typically needed for giant planet formation to be efficient \citep[c.f.][]{2015Bitsch,Emsenhuber2021}.

\bsp	
\label{lastpage}
\end{document}